\documentclass[a4paper,11pt]{article}
\pdfoutput=1 % if your are submitting a pdflatex (i.e. if you have
\usepackage{jcappub} % for details on the use of the package, please
\usepackage[english]{babel}

\newcommand{\w}{\omega}
\newcommand{\lam}{\lambda}

\title{\huge New effects of non-standard self-interactions\\ of neutrinos in a supernova}

\author{Anirban Das,}
\author{Amol Dighe}
\author{and Manibrata Sen}
\affiliation{Tata Institute of Fundamental Research,
             Homi Bhabha Road, Mumbai, 400005, India.}

\emailAdd{anirbandas@theory.tifr.res.in}
\emailAdd{amol@theory.tifr.res.in}
\emailAdd{manibrata@theory.tifr.res.in }

\abstract{
Neutrino self-interactions are known to lead to non-linear collective flavor \mbox{oscillations} in a core-collapse supernova.
We point out new possible effects of non-standard self-interactions (NSSI) of neutrinos on  flavor conversions in a two-flavor framework. 
 We show that, for a single-energy neutrino-antineutrino ensemble, a flavor 
instability is generated even in normal hierarchy for large enough NSSI.
Using a toy model for the neutrino spectra, we show that flavor-preserving NSSI lead to pinching of spectral swaps, while flavor-violating NSSI cause swaps to develop away from a spectral crossing or even in the absence of a spectral crossing. Consequently, NSSI could give rise to collective oscillations and spectral splits even
during neutronization burst, for both hierarchies.
}


\begin{document}
\hfill { TIFR/TH/17-18}
\maketitle
\flushbottom

\section{Introduction}
\label{sec:1}
It has been known for quite some time that neutrino-neutrino interactions in a dense gas of neutrinos and antineutrinos can lead to non-linear 
collective flavor conversions~\cite{Duan:2005cp,Duan:2006an,Duan:2006jv}.
Deep inside a supernova (SN) core, the neutrino density is so high that neutrinos/antineutrinos streaming out experience a potential not only due to the ordinary matter 
background, but also due to the ambient neutrinos~\cite{Pantaleone:1994ns,Kostelecky:1994dt,Samuel:1995ri}. This
makes the problem non-linear, and new consequences emerge. In~\cite{Hannestad:2006nj}, two stages of collective effects were recognized: (i) ``synchronized''
oscillations, where neutrinos of all energies oscillate with a single frequency but small amplitude---due to the small mixing angle in matter ~\cite{Wong:2002sc}, and 
(ii) ``bipolar'' oscillations, where even the small mixing angle
can lead to a complete flavor conversion for inverted mass hierarchy (IH). In the case of normal mass hierarchy (NH), no bipolar oscillations take place \cite{Duan:2005cp}. This analysis was restricted to two flavors: 
electron neutrino $(\nu_e)$ and $\alpha$-flavor neutrino $(\nu_\alpha)$, where $\alpha=\mu$, $\tau,$ or any linear combination of both. Bipolar oscillations were shown to be equivalent to an inverted pendulum in flavor space, 
where swinging of the pendulum from an unstable inverted position was shown to be equivalent to flavor oscillations.

The bipolar oscillations further culminate in spectral splits. For a single spectral split, all $\nu_e$s beyond a certain critical energy
would convert to $\nu_\alpha$, while below this energy, the $\nu_e$s would emerge in their original flavor~\cite{Duan:2006an}. An analytical understanding of the phenomenon was offered ~\cite{Raffelt:2007xt,Raffelt:2007cb} using a simple box spectrum in the variable $\w\equiv\pm\Delta m^2/(2E)$, where the positive (negative) sign stands for neutrinos (antineutrinos).
Multiple spectral splits~\cite{Fogli:2007bk,Fogli:2008pt} were explained analytically~\cite{Dasgupta:2009mg} in terms of the development of spectral swaps in $g_\w$, where
\begin{eqnarray}\label{spectrum}
g_\w &\propto& F_{\nu_e}(\w)-F_{\nu_\alpha}(\w) \qquad{\rm for~ }\w>0 \; ,\nonumber\\
     &\propto& F_{\bar{\nu}_\alpha}(\w)-F_{\bar{\nu}_e}(\w)\qquad {\rm for~}\w<0\;.
\end{eqnarray}
It was shown that in IH (NH), any positive (negative) crossing of the $g_\w$-spectra is unstable, thereby causing bipolar oscillations leading to a spectral swap. Each edge of the spectral swap corresponds to a spectral split.
The widths of the spectral swap on the two sides of the crossing are determined by
the development of bipolar oscillations, however they are always related by flavor lepton-number conservation. 
Although the above analyses were carried out in the two-flavor framework with a single angle approximation~\cite{Duan:2006an}, important features of spectral splits were brought out even in such simplified scenarios. 

The three-flavor mixing effects can be largely understood in terms of 
stepwise two flavor effects~\cite{Dasgupta:2008cd,Dasgupta:2007ws,Friedland:2010sc,Choubey:2010up}.
Multi-angle effects \cite{Fogli:2007bk} were shown to cause smearing of features in bipolar oscillations, leading to smoothening of spectral splits.
However, such multi-angle decoherence can be suppressed for sufficient neutrino-antineutrino asymmetry, which can exist in the deleptonisation flux in a realistic supernova~\cite{EstebanPretel:2007ec}.
The inclusion of multi-angle effects can  also result in the suppression of bipolar oscillations in presence of matter~\cite{Duan:2006an,Fogli:2007bk,EstebanPretel:2008ni,Sawyer:2008zs,Chakraborty:2011gd,Sarikas:2012ad}.

We are still far from having a complete analytical picture of the flavor instabilities that lead to spectral swaps. However, one can get an understanding of the onset of these instabilities using a linearized stability analysis~\cite{Banerjee:2011fj}. For an isotropic gas of neutrinos and antineutrinos, the rate of development of these instabilities is proportional to $\sqrt{\w\mu}$, where $\mu$ is the rate of interaction with ambient neutrinos~\cite{Chakraborty:2015tfa}.
If the assumptions of homogeneity and isotropy do not hold, then new instabilities can arise even deeper inside the star, which would result in faster flavor conversions  with a rate proportional to $\mu$~\cite{Sawyer:2015dsa}.
A necessary condition for development of these fast oscillations is a non-trivial angular distribution
among the different neutrino species, essentially, a crossing in the neutrino angular spectra~\cite{Chakraborty:2016lct,Dasgupta:2016dbv,Sen:2017ogt}. A new approach to understand this, using dispersion relations, was developed in~\cite{Izaguirre:2016gsx}.

All the above analyses have been performed within the context of the Standard Model (SM). However, extensions to the SM can predict new non-standard self-interactions (NSSI) of neutrinos.
While non-standard interactions of neutrinos with charged fermions are well- constrained~\cite{Bergmann:1999pk,Bergmann:2000gp,Gago:2001si,Friedland:2004pp,Guzzo:2004ue,EstebanPretel:2007yu}, NSSI
are very loosely constrained and can be as large as in the SM, if not larger~\cite{Bilenky:1992xn,Bilenky:1994ma,Masso:1994ww,Bilenky:1999dn}. This is primarily because neutrino-neutrino interactions have not been directly observed yet, and hence 
it is very difficult to put bounds on them. The NSSI would give rise to an effective operator of the form 
$G_F \left(G^{\alpha\beta}\, \bar{\nu}_{{\rm L}\alpha}\gamma^\mu\nu_{{\rm L}\beta}\right)\,\left(G^{\zeta\eta}\,\bar{\nu}_{{\rm L}\zeta}\gamma_\mu\nu_{{\rm L}\eta}\right)$, 
where the coupling matrix $G$ contains both standard and non-standard components. When $\alpha=\beta$, the coupling $G^{\alpha\beta}$ is flavor-preserving and we call such interactions as flavor-preserving NSSI (FP-NSSI). Similarly, 
when $\alpha\neq\beta$, then $G^{\alpha\beta}$ would be flavor-violating and we refer to such interactions flavor-violating NSSI (FV-NSSI). 
Such NSSI
of neutrinos within a core-collapse SN can have important impact on 
collective oscillations.

The framework for analyzing the effect of NSSI on collective oscillations was first developed in~\cite{Blennow:2008er},
which showed that FV-NSSI can cause complete flavor conversions even in the absence of any mixing.
Motivated by this observation and the rather loose constraints on NSSI, in this paper we perform a detailed study of the effect of NSSI on supernova neutrino flavor evolution. We find that the presence of NSSI makes us reconsider many of the popular notions
of flavor evolutions of dense neutrino streams. 
Below we briefly outline these notions, before 
going into a more rigorous analysis in the following sections.
\begin{itemize}
 \item In a completely spherically symmetric ensemble of neutrinos and antineutrinos of a single energy, flavor conversions are known to happen only in IH. In the case of NH, only a breaking of these initial spherical 
 symmetries can lead to conversions \cite{Raffelt:2013rqa}. 
      
       We will find that presence of NSSI couplings larger than the SM couplings can lead to conversions in NH, without the need to break any of the symmetries of the initial setup.
  In the flavor-pendulum language, the NSSI act like an external force which can overturn the stable position of the pendulum, thereby making it unstable. Thus with a large enough NSSI, an NH scenario can
  mimic a standard IH scenario, and vice versa.

  \item Spectral swaps develop around the zero crossing of the $g_\w$ spectra \cite{Dasgupta:2009mg}. 
  
        FV-NSSI violates flavor lepton number, which may cause the swap to develop away from the zero crossing of the $g_\w$ spectra. 
        A spectral crossing is then no longer necessary for the development of swaps.
        This could give rise to collective oscillations during the neutronization epoch, which are absent otherwise. 
        Distinct splits in the neutrino spectra during this epoch can then be a signal of NSSI.
\end{itemize}

We expand on these ideas in the following sections. In Section~\ref{sec:2}, we set up the formalism and analyze the evolution for a system of $\nu$ and $\bar{\nu}$ of a single energy
in presence of NSSI and demonstrate collective effects in NH for large enough NSSI.

In Section~\ref{sec:3}, we consider a simple asymmetric box spectrum and illustrate the effects of NSSI parameters analytically as well as numerically. In Section~\ref{sec:4}, we highlight the presence of spectral splits during
neutronization epoch as a result of NSSI. Finally, in Section~\ref{sec:5}, we discuss our results and conclude.

%%%%%%%%%%%%%%%%%%%%%%%%%%%%%%%%%%%%%%%%%%%%%%%%%%%%%%%%%%%%%%%%%%%%%%%%%%%%%%%%%%%%%%%%%%%%%%%%%%%%%%%%%%%%%%%%%%%%%%%%%%%%%%%%%%%%%%%%%%%%%%%%%%%%%%%%%%%%%%%%%%%%%%%%%%%%%%%%%%%%%%%%

\section{Flavour evolution in presence of NSSI}
\label{sec:2}

We consider a two-flavor setup, consisting of $\nu_e$ and $\nu_\alpha$ in the presence of NSSI. We write the entire formalism 
in terms of momentum-dependent $2\times2$ density matrices $\varrho_\mathbf{p}$, where the dependence on position $x$ and time $t$ is implicit. In terms of these, the equation of motion (EoM) for each mode ${\bf p}$ is given by \cite{Sigl:1992fn,Strack:2005ux}
\begin{equation}\label{eq:eom}
\partial_t \varrho_{{\bf p}} + {\bf v}_{\bf p} \cdot \nabla_{\bf x} \varrho_{{\bf p}}=- i [\Omega_{{\bf p}}, \varrho_{{\bf p}}] +{\mathcal C}[ \varrho_{{\bf p}}]
\,\ .
\end{equation}
The Hamiltonian matrix $\Omega_{{\bf p}}$,
%...............................................................
\begin{equation}
\Omega_{{\bf p}}= \Omega^{{\rm vac}}_{\bf p} + \Omega^{\rm MSW} + \Omega^{\nu\nu}_{\bf p}\,\ ,
\label{eq:ham}
\end{equation}
%............................................................
contains the vacuum, matter and self-interaction terms.
 
The matrix of vacuum oscillation frequency in the mass basis is 
\begin{equation}
 \Omega^{{\rm vac}}_{\bf p}= \textrm{diag}(m_1^2, m_2^2)/(2|{\bf p}|)\, ,
\end{equation}
where $E=|{\bf p}|$ for 
ultra-relativistic neutrinos.  For the antineutrino density matrix $\bar{\varrho}_{\bf p}$, the same EoMs apply but with the replacement, $\Omega^{{\rm vac}} \to - \Omega^{{\rm vac}}$; 
thus antineutrinos of energy $E$ can be thought of as neutrinos of energy $-E$,
having identical EoMs . The charged-current Mikheev-Smirnov-Wolfenstein (MSW) potential term in Eq.~(\ref{eq:eom}), due to the background 
electron density $n_e$, is represented by
%...................................................................................
\begin{equation}
\Omega^{\rm MSW}=  \lambda\,\ \textrm{diag} (1,0) \,\ 
\end{equation}
%......................................................................
in the weak interaction basis, where $\lambda \equiv\sqrt{2} G_F n_e$. 
The most general form of the effective Hamiltonian due to self interactions is given by
%........................................................
\begin{equation}
\Omega^{\nu\nu}_{\bf p} = \sqrt{2} G_F \int \frac{d^3 {\bf q}}{(2 \pi)^3}  (1 -{\bf v}_{\bf p}\cdot {\bf v}_{\bf q})\left\{G({\varrho_{\bf q}} - {\bar\varrho_{\bf q}})G + G~{\rm Tr}\left[({\varrho_{\bf q}} - {\bar\varrho_{\bf q}}) G\right] \right\} \,\ ,
\end{equation}
where the term $(1 -{\bf v}_{\bf p}\cdot {\bf v}_{\bf q})$ leads to multi-angle effects due to neutrinos moving on different trajectories.
In the SM, the dimensionless coupling matrix $G$ is an identity matrix. After including NSSI, the most general coupling matrix is 
given by 
\begin{equation}\label{coupling}
 G=\begin{bmatrix}
      1+\gamma_{ee} & \gamma_{ex} \\
      \gamma_{ex}^* & 1+\gamma_{xx}
      \end{bmatrix}.
\end{equation}
The bounds on $\gamma_{\alpha\beta}$ are very weak since processes involving neutrino self-interactions are rare and difficult to observe.
Loose bounds on these four-neutrino contact interactions can be put from low-energy $\pi^+,\,K^+$ decays \cite{Bardin:1970wq} and from SN1987A data \cite{Kolb:1987qy}.
However, much stronger constraints on neutrino NSSI come from LEP data. The presence of non-standard neutrino coupling can give rise to a new decay channel $Z\rightarrow \nu\bar{\nu} \rightarrow \nu\bar{\nu}\nu\bar{\nu}$, 
which modifies the invisible $Z$-width predicted by the SM \cite{Bilenky:1992xn}. 
Alternatively, such new interactions can contribute to loop corrections in a SM process, for example, in the  $Z\rightarrow\nu\nu$ decay channel.
The invisible $Z$-width is measured with an accuracy better than $1\%$ and this can put more stringent bounds on the coupling\,\cite{Bilenky:1994ma,Bilenky:1999dn}. Assuming the NSSI are due to the presence of a new gauge boson, these bounds directly translate into $ |\gamma_{ee}|,\, |\gamma_{xx}|\,{\rm and}\,|\gamma_{ex}|\sim\, \mathcal{O}(1)$. Stronger bounds can come from primordial nucleosynthesis, however one needs to assume the presence of right-handed neutrinos \cite{Masso:1994ww}.

The last term on RHS in Eq.\,(\ref{eq:eom}) represents collisions between different neutrino flavors. For simplicity, we neglect this term in our analysis.
 Following the notation in~\cite{Blennow:2008er}, we write all matrices in Pauli basis as follows:
%........................................................
\begin{eqnarray}\label{pauli}
 \Omega^{{\rm vac}}_{\bf p}	 &=&\frac{1}{2}\left(\omega_0 {\mathbb I}+\omega_{\bf p}{\bf B}\cdot {\bf \sigma}\right),\qquad\Omega^{\rm MSW}=\frac{1}{2}\left(\lambda{\mathbb I}+\lambda{\bf L}\cdot {\bf\sigma}\right),\nonumber\\
 \varrho_{\bf p}                 &=&\frac{1}{2}\left(f_{\bf p} {\mathbb I}+n_{\bar{\nu}}{\bf P_{\bf p}}\cdot {\bf\sigma}\right),\qquad
 \overline{\varrho}_{\bf p}  =\frac{1}{2}\left(\bar{f_{\bf p}} {\mathbb I}+n_{\bar{\nu}}\overline{{\bf P}}_{\bf p}\cdot {\bf\sigma}\right),
\end{eqnarray}
where $\mathbf{P}_\mathbf{p}$ and $\overline{\mathbf{P}}_\mathbf{p}$ may be interpreted as the polarization vectors for neutrinos and anti-neutrinos,  respectively.
The coordinate system is chosen such that the polarization vector pointing in the $+z$ indicates $\nu_e$ whereas that in the $-z$ direction indicates $\nu_\alpha$.
The neutrino density is given by integrating the momentum distribution function $f_\mathbf{p}$ over all momentum modes,
$n_\nu \equiv \int d^3\mathbf{p}\, f_\mathbf{p}$ and $n_{\bar{\nu}} \equiv \int d^3\mathbf{p}\, \bar{f}_\mathbf{p}$. The normalization, thus, is such that
$|\overline{\bf P}_{\bf p}|=1$.
The vacuum Hamiltonian may be interpreted as an external magnetic field, given by $\mathbf{B} = (\sin2\vartheta_0, 0, -\cos2\vartheta_0)$, where $\vartheta_0$ is the vacuum mixing angle between the two flavors. 
The MSW potential $\Omega^{\rm MSW}$ is characterized by its magnitude $\lambda$ and a unit vector $ \mathbf{L} = (0, 0, 1)$ . 

The coupling matrix $G$ is similarly represented as 
\begin{equation}\label{GG}
G=\frac{1}{2}\left(g_0{\mathbb I}+{\boldsymbol g}\cdot {\bf\sigma}\right) \,.
\end{equation}
The four-vector  $g = \{g_0, {\boldsymbol g}\}$
is a measure of the net neutrino-neutrino coupling. The SM corresponds to $g_0=2$ and $|{\boldsymbol g}|=0$.
Thus, the vector ${\boldsymbol g}$ indicates the strength of NSSI. Clearly, $g_0=2+\gamma_{ee}+\gamma_{xx},\,g_1=2\,{\rm Re}(\gamma_{ex}),\,g_2=2\,{\rm Im}(\gamma_{ex}^*)$ and $g_3=\gamma_{ee}-\gamma_{xx}$. Thus, $g_0$ and $g_3$ are flavor-preserving NSSI (FP-NSSI) couplings while $g_1$ and $g_2$ are flavor-violating NSSI (FV-NSSI) couplings. Note that we can redefine the phase of $\nu_\alpha$ such that $g_2=0$. The NSSI can thus be parameterized by $g_0,\,g_1$ and $g_3$.

With the definitions in Eq.\,(\ref{pauli}) and Eq.\,(\ref{GG}), the EoMs take the  form
%........................................................
\begin{eqnarray}\label{eq:eom1}
 \dot{{\bf P}}_{\bf p}      &=&\left(\omega_{\bf p} {\bf B}+ \lambda{\bf L} + \Omega^{\nu\nu}_{\bf p} \right) \times {\bf P_p}, \nonumber\\
 \dot{\overline{{\bf P}}}_{\bf p}&=&\left(-\omega_{\bf p} {\bf B}+ \lambda{\bf L} + \Omega^{\nu\nu}_{\bf p} \right) \times \overline{{\bf P}}_{\bf p},
 \end{eqnarray}
%........................................................
where
%........................................................
\begin{equation}\label{eq:nu-int}
\Omega^{\nu\nu}_{\bf p} = \mu \int \frac{d^3 {\bf q}}{(2 \pi)^3} (1 -{\bf v}_{\bf p}\cdot {\bf v}_{\bf q})\left\{\frac{1}{4}\left(g_0^2-|{\boldsymbol g}|^2\right)\left({\bf P_q-\overline{P}_q}\right)+\bigl[g_o\xi+{\boldsymbol g}\cdot\left({\bf P_q-\overline{P}_q}\right)\bigr]{\boldsymbol g}\right\} \,,
\end{equation}
%..........................................................
and  $\mu\equiv\sqrt{2} G_F n_{\bar{\nu}}$. The variable $\xi$ parameterizes the neutrino-antineutrino asymmetry such that $n_{\nu} = (1+\xi)n_{\bar{\nu}}$. We further note that the parameter $g_0$ can be scaled away by the following redefinition in Eq.\,(\ref{eq:nu-int}) :
\begin{equation}\label{rescale}
 \mu \to \mu (g_0/2)^2\,, \qquad {\boldsymbol g} \to {\boldsymbol g}/(g_0/2).
\end{equation}
This yields
\begin{equation}\label{eq:nu-int2}
 \Omega^{\nu\nu}_{\bf p} = \mu \int \frac{d^3 {\bf q}}{(2 \pi)^3} (1 -{\bf v}_{\bf p}\cdot {\bf v}_{\bf q})\left\{\left(1-\frac{|{\boldsymbol g}|^2}{4}\right)\left({\bf P_q-\overline{P}_q}\right)+\bigl[2\xi+{\boldsymbol g}\cdot\left({\bf P_q-\overline{P}_q}\right)\bigr]{\boldsymbol g}\right\} \,.
\end{equation}
Henceforth, we will work in terms of the rescaled ${\boldsymbol g}$ and $\Omega^{\nu\nu}_{\bf p}$ .

It is interesting to note that $\Omega^{\nu\nu}_{\bf p}$  in Eq.\,(\ref{eq:nu-int2}) has two types of terms: the first one is the SM neutrino-neutrino interaction term modified due to $|{\boldsymbol g}|$. 
If this were the only term present, the EoMs would represent a precessing top with a modified $\mu$ given by $\mu\rightarrow\mu\left(1-|{\boldsymbol g}|^2/4\right)$. 
However, there is a subtle difference. The modified $\mu$ term can now change sign depending on the value of $|{\boldsymbol g}|$ and affect the motion of the top.
The second term gives rise to a term  of the form $\chi(t){\boldsymbol g}\times {\bf P}$(for neutrinos) or $\chi(t){\boldsymbol g}\times \overline{{\bf P}}$ (for antineutrinos) in the EoMs [Eq.\,(\ref{eq:eom1})] . It represents 
the equation of a precessing top around the direction ${\boldsymbol g}$ with a time-dependent frequency $\chi(t)$.
This allows us to interpret the NSSI vector as an external force on the precessing top.

In the further analysis, we work with a single-angle approximation, where the problem has a spherical symmetry and all neutrinos are emitted with the same ``emission angle''~\cite{Duan:2006an}.
In this scenario, the factor of $(1 -{\bf v}_{\bf p}\cdot {\bf v}_{\bf q})$ drops out of the integral.

\subsection{The flavor pendulum}
To understand the flavor evolution more clearly, we follow the analysis in~\cite{Hannestad:2006nj}, and 
rewrite our EoMs for a single momentum mode ${\bf p}$ in terms of the new vectors ${\bf D}\equiv{\bf P-\overline{P}}$,\,\, ${\bf S}\equiv{\bf P+\overline{P}} $.
The EoMs are exactly identical with
\begin{equation}\label{redef}
 \mu\rightarrow\widetilde{\mu}\equiv\mu\left(1-|{\boldsymbol g}|^2/4\right)\,,\qquad \lam{\bf L}\rightarrow\widetilde{\lambda}\widetilde{{\bf L}}\equiv\lambda{\bf L}+ \mu \left(2\xi + {\bf D}\cdot{\boldsymbol g}\right){\boldsymbol g}\,,
\end{equation}
where the vector $\widetilde{\bf L}$ is defined such that it is normalized to unity. 
In terms of the vector
\begin{equation}\label{Qvec}
 {\bf Q}\equiv{\bf S}-(\omega/\widetilde{\mu}){\bf B}
\end{equation}
 the EoMs are
\begin{eqnarray}\label{eq:eom2a}
 \dot{{\bf Q}}      &=&\widetilde{\mu}\, {\bf D}\times {\bf Q} +\widetilde{\lambda}\, \widetilde{{\bf L}}\times {\bf S}  \nonumber\\
\dot{{\bf D}}       &=&\omega\, {\bf B}\times {\bf Q}+\widetilde{\lambda}\, \widetilde{{\bf L}}\times {\bf D}\, .
 \end{eqnarray}
We observe that in the scenario $\widetilde{\lambda}=0$, $|{\bf Q}|$ is conserved. Thus ${\bf Q}$
describes a spherical pendulum in the flavor space with length $|{\bf Q}|$, as in \cite{Hannestad:2006nj}. In the subsequent subsections, we study the motion of this pendulum 
for a constant as well as a realistic, decreasing neutrino density profile.

\subsection{Constant neutrino-neutrino potential}
\subsubsection{Small mixing angle}
\label{sma}
\begin{figure}[!t]
\begin{center}
\includegraphics[width=0.48\textwidth]{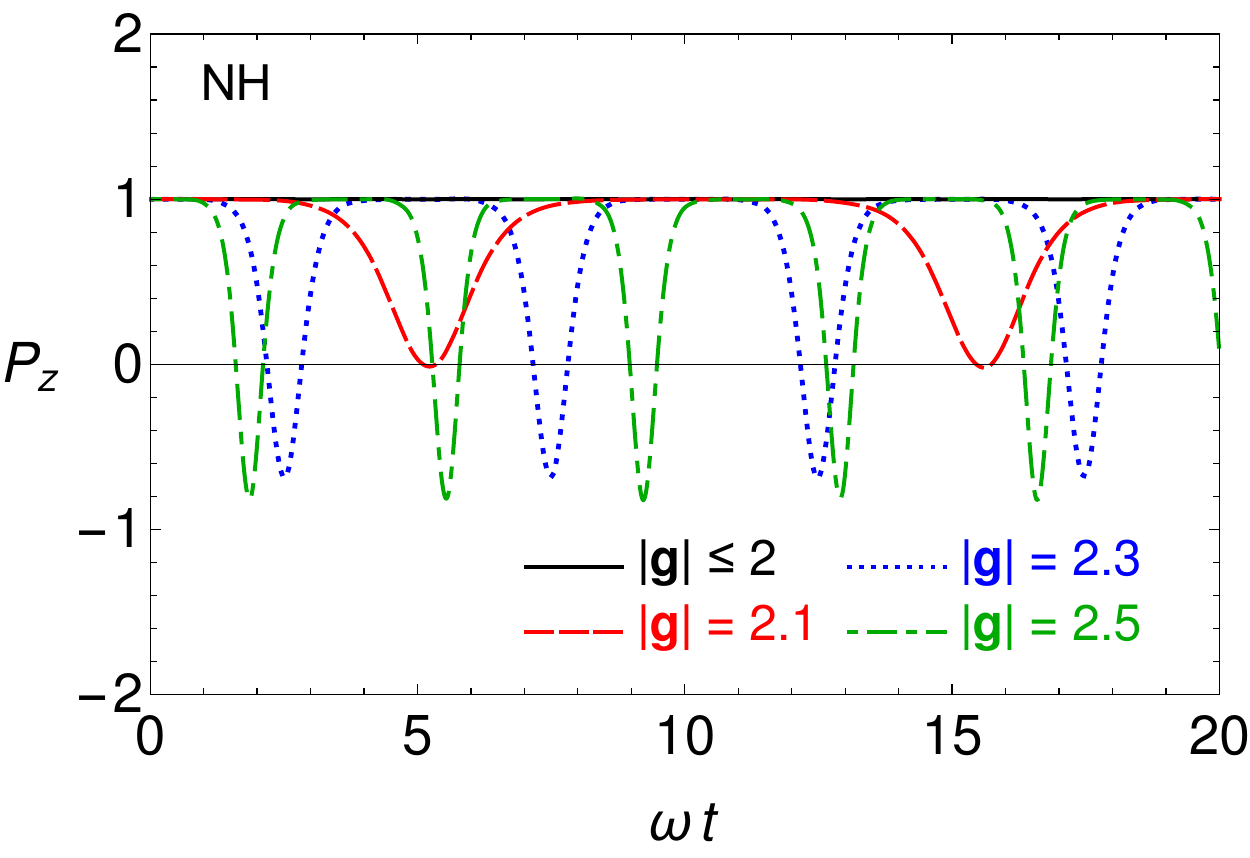}~~\includegraphics[width=0.48\textwidth]{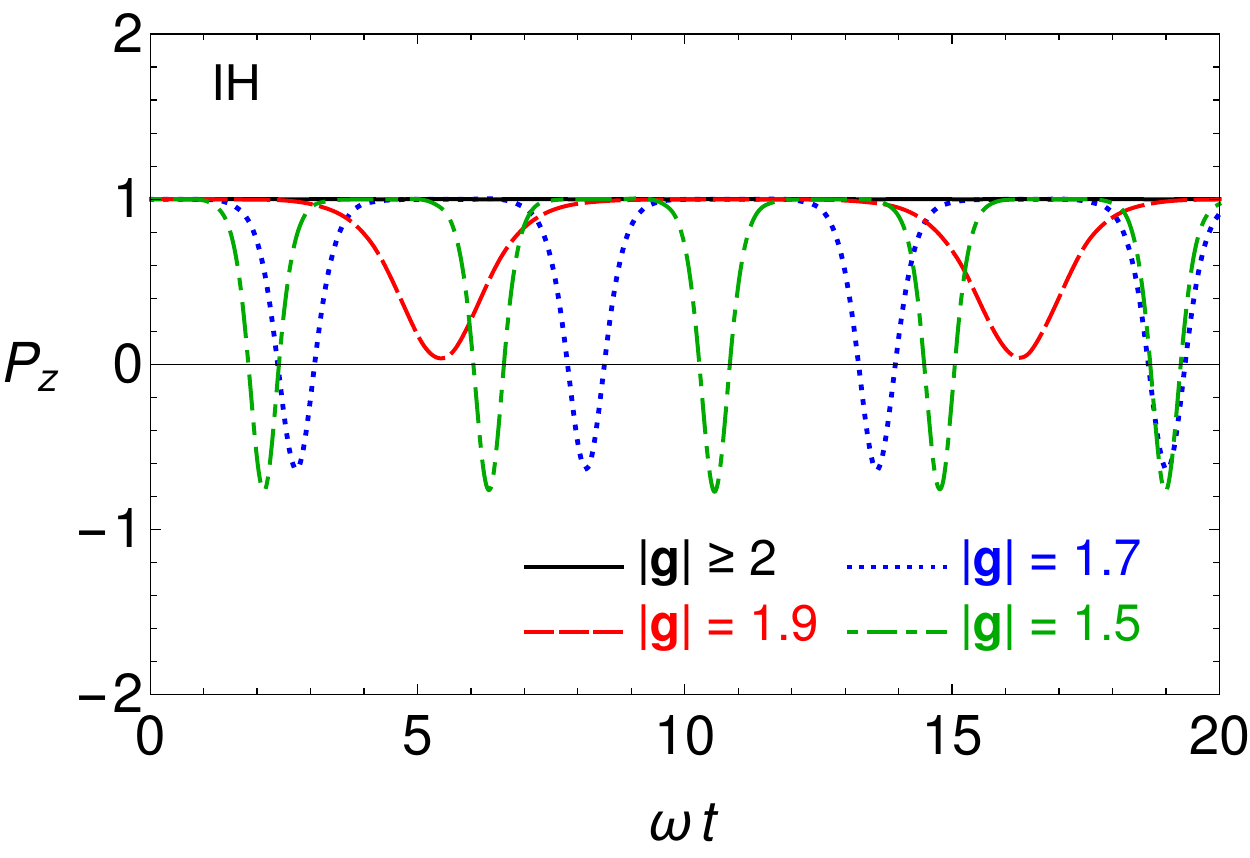}
\caption{Time evolution of $P_z$ and $\overline{P}_z$ for four different values of $|{\boldsymbol g}|$, for NH (left) and IH (right). The other parameters are $\mu=10\,\w$, $\lam=0$, and $\vartheta_0=0.01$ .}
\label{fig1}
\end{center}
\end{figure}

In this section, we confine ourselves to a fixed neutrino density and study the flavor evolution of the system. 
In Fig.\,\ref{fig1}, we show the variation of $P_z$, i.e., $z$-component of the polarization vector ${\bf P}$, with time for three different values of $|{\boldsymbol g}|$.
We consider zero asymmetry and for simplicity, put $g_1,~g_2=0$, hence $|{\boldsymbol g}|=g_3$. Note that ${\bf D}\cdot{\boldsymbol g}=D_z g_3$ is almost conserved in this case, as the mixing angle $\vartheta_0$ is small. Since $D_z(0)=0$, we have ${\bf D}\cdot{\boldsymbol g}\simeq0$ and hence $\widetilde{\lambda}\simeq\lam$. We choose the matter potential $\lam=0$ to start with.

In NH, there is no instability in the SM, i.e., $|{\boldsymbol g}|=0$ , since this is similar to a normal pendulum. With NSSI, since $\xi=0$ and ${\bf D}\cdot{\boldsymbol g}\simeq0$, the second term in Eq.\,(\ref{eq:nu-int2}) vanishes and $\mu$ is simply modified to $\widetilde{\mu}$. The stability of the pendulum holds for $|{\boldsymbol g}|\leq\,2$. However, if $|{\boldsymbol g}|>\,2$, i.e., when the sign of $\widetilde{\mu}$ reverses, collective oscillations start taking place.
This can be clearly seen in the left panel of Fig.\,\ref{fig1}.
As $|{\boldsymbol g}|$ is increased further, the frequency of collective oscillations also increases.
The results in IH are complementary; for $0\,\leq|{\boldsymbol g}|\leq\, 2$, we observe an instability, whereas for $|{\boldsymbol g}|>\,2$, the instability vanishes (Fig.\,\ref{fig1}, right panel).

The features observed in Fig.\,\ref{fig1} may be understood as follows.
In the absence of $\lam$, Eq.\,(\ref{eq:eom2a}) becomes,
\begin{eqnarray}\label{eq:eom2b}
 \dot{{\bf Q}}      &=&\widetilde{\mu}\, {\bf D}\times {\bf Q}   \nonumber\\
\dot{{\bf D}}       &=&\omega\, {\bf B}\times {\bf Q}\, .
 \end{eqnarray}
Clearly ${\bf B}\cdot{\bf D}$ is conserved 
. Moreover,  With initial conditions ${\bf P}(0)=\overline{{\bf P}}(0)=\left(0,0,1\right)$, the scenario is equivalent 
to that of a pendulum  whose oscillations are 
confined to a plane defined by ${\bf B}$ and z-axis. In terms of ${\bf Q}=|{\bf Q}|\left(\sin\varphi,0,\cos\varphi\right)$, we have 
\begin{equation}\label{eq:eom3}
 \ddot{\varphi}=-\omega\widetilde{\mu} |{\bf Q}| \sin(\varphi+2\vartheta_0)\,.
\end{equation}
For small $\vartheta_0$ and $\varphi$, Eq.\,(\ref{eq:eom3}) corresponds to a harmonic oscillator for $\w\widetilde{\mu}>0$ and 
an inverted pendulum for $\w\widetilde{\mu}<0$. This is the reason why we observe collective oscillations in NH for $|{\boldsymbol g}| >\,2$ when the sign of $\w\widetilde{\mu}$ becomes negative and we enter the inverted pendulum phase.
Note that here the NSSI themselves have made the pendulum unstable, without 
breaking the spherical symmetry of the system. On the other hand, in IH, when $|{\boldsymbol g}| >\,2$ , the sign of $\w\widetilde{\mu} $ becomes positive and collective oscillations are suppressed.

 Using the initial conditions $\varphi(0)\simeq-\left(\w/\widetilde{\mu}|{\bf Q}|\right)2\vartheta_0$ and $\dot{\varphi}(0)=0$, the solution to the above inverted pendulum $(\w\widetilde{\mu}<0)$ for small $\vartheta_0$ and $\varphi$ is
 \begin{equation}
  \varphi(t)=2\vartheta_0\left[1-\left(1+\frac{\w}{\widetilde{\mu}|{\bf Q}|}\right)\cosh\left(\sqrt{\left|\omega\widetilde{\mu} {\bf Q}\right| }\, t\right)\right]\,.
 \end{equation} 
 During the pendular oscillations, the time taken for $\varphi(t)$ to become of order unity is then 
\begin{equation}
 \tau\simeq
 \left|\frac{1}{\left|\omega\widetilde{\mu} {\bf Q}\right|} \ln\left[\vartheta_0\left(1+\frac{\w}{\widetilde{\mu}|{\bf Q}|}\right)\right]\right| \,.
\end{equation}
Thus, the frequency of collective oscillations is larger when $\w\widetilde{\mu}$ becomes more negative. This corresponds to
$|{\boldsymbol g}|$ increasing beyond $2$ in NH and $|{\boldsymbol g}|$ decreasing below $2$ in IH, as observed in Fig.\,\ref{fig1}.

Now that we have demonstrated that instabilities in NH for $|{\boldsymbol g}|>\,2$ are similar to 
IH for $|{\boldsymbol g}|<\,2$, for further analysis we will confine ourselves to the scenario with IH and $|{\boldsymbol g}|<\,2$\,.

\begin{figure}[!t]
\begin{center}
\includegraphics[width=0.48\textwidth]{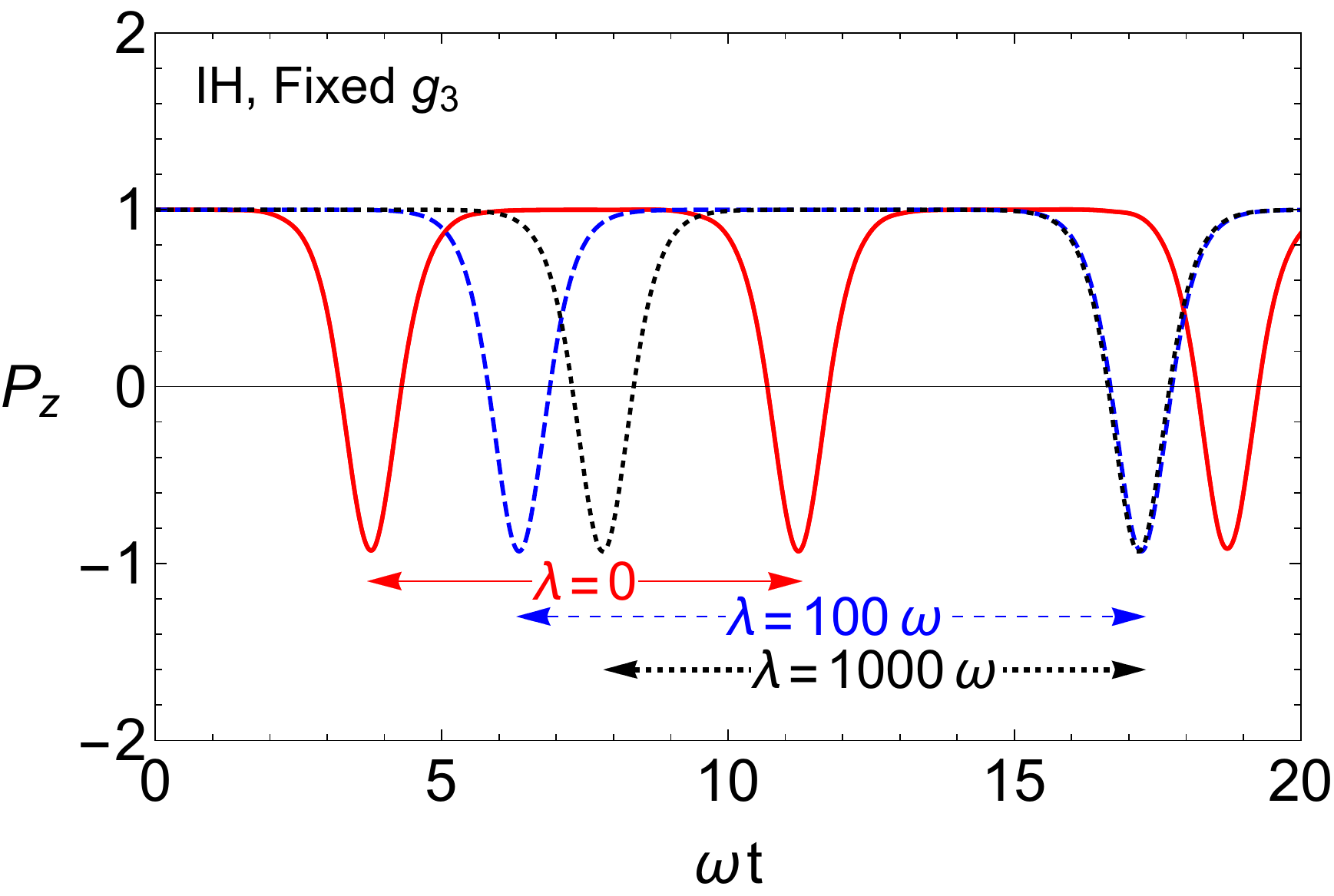}~~\includegraphics[width=0.485\textwidth]{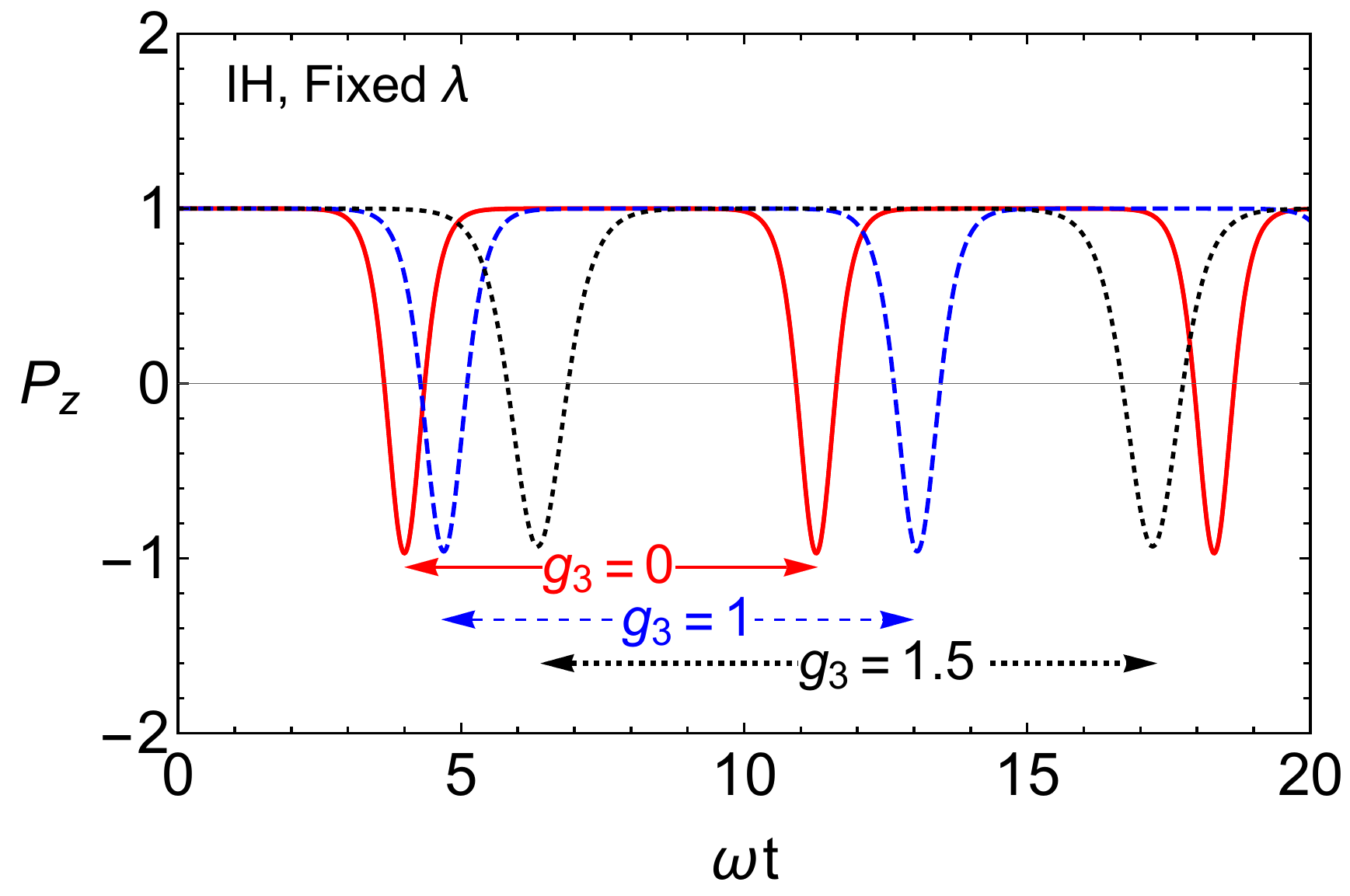}
\caption{Left: Time evolution of $P_z$ in IH with three different values of $\lam=0,~100\,\w,~1000\,\w$ for $g_3=1.5$.
Right: Time evolution of $P_z$ in IH with three different values of $g_3=0,~1,~1.5$ for $\lam=100\,\w.$}
\label{fig2}
\end{center}
\end{figure}

Next we will show  the effect of a non-zero $\widetilde{\lam}=\lam$. This term is identical for both neutrinos and antineutrinos, and hence can be rotated away
in a co-rotating frame, in the single-angle approximation, as shown in \cite{Hannestad:2006nj}. 
 Now ${\bf B}\cdot{\bf D}$ and $|{\bf Q}|$ are not exactly conserved, but 
 exhibit fluctuations with a frequency of $\lam$. These can average out to zero for large $\lam$.
 We  plot the time evolution of $P_z$ with finite matter effect in Fig.\,\ref{fig2} (left panel).  It can be seen that the presence of matter effects does not change the qualitative nature of the plots, however, the value of  $\tau$ increases with $\lam$ as has already been noticed \cite{Hannestad:2006nj}. The right panel of Fig.\,\ref{fig2}
shows the effect of changing $g_3$ in presence of a fixed matter density $\lam$. We observe that changing $g_3$ also gives qualitatively similar results and leads to extension $\tau$.
Thus, a non-zero $g_3$ acts like an extra matter term in the system. This can also be discerned from Eq.\,\ref{redef}.

\subsubsection{Large mixing angle}

\begin{figure}[!t]
\centering
 \includegraphics[width=1.02\textwidth]{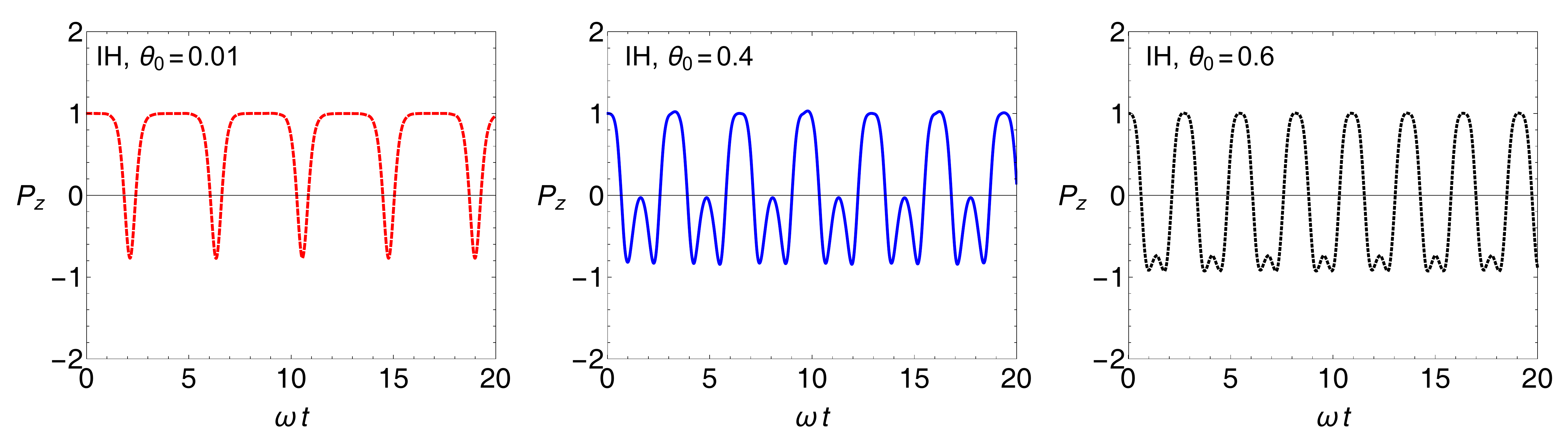}
\caption{Time evolution of $P_z$ and $\overline{P}_z$ in IH for three different values of $\vartheta_0 = 0.01,0.4,0.6 $. We take $g_3=1.5$ and $\mu = 10\,\w$.}
\label{fig1a}
\end{figure}
\begin{figure}[!ht]
\centering
 \includegraphics[scale=0.29]{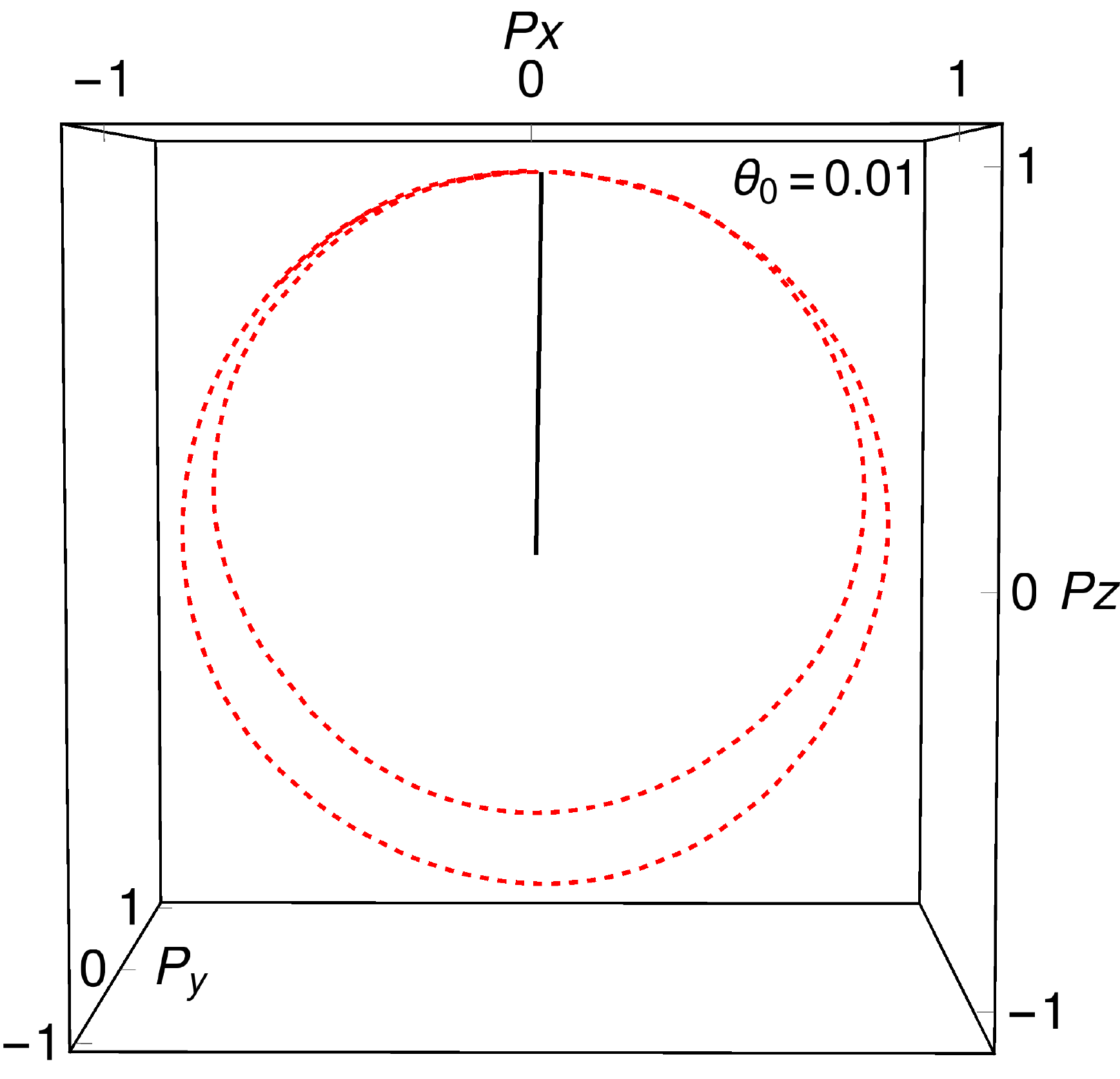}\includegraphics[scale=0.29]{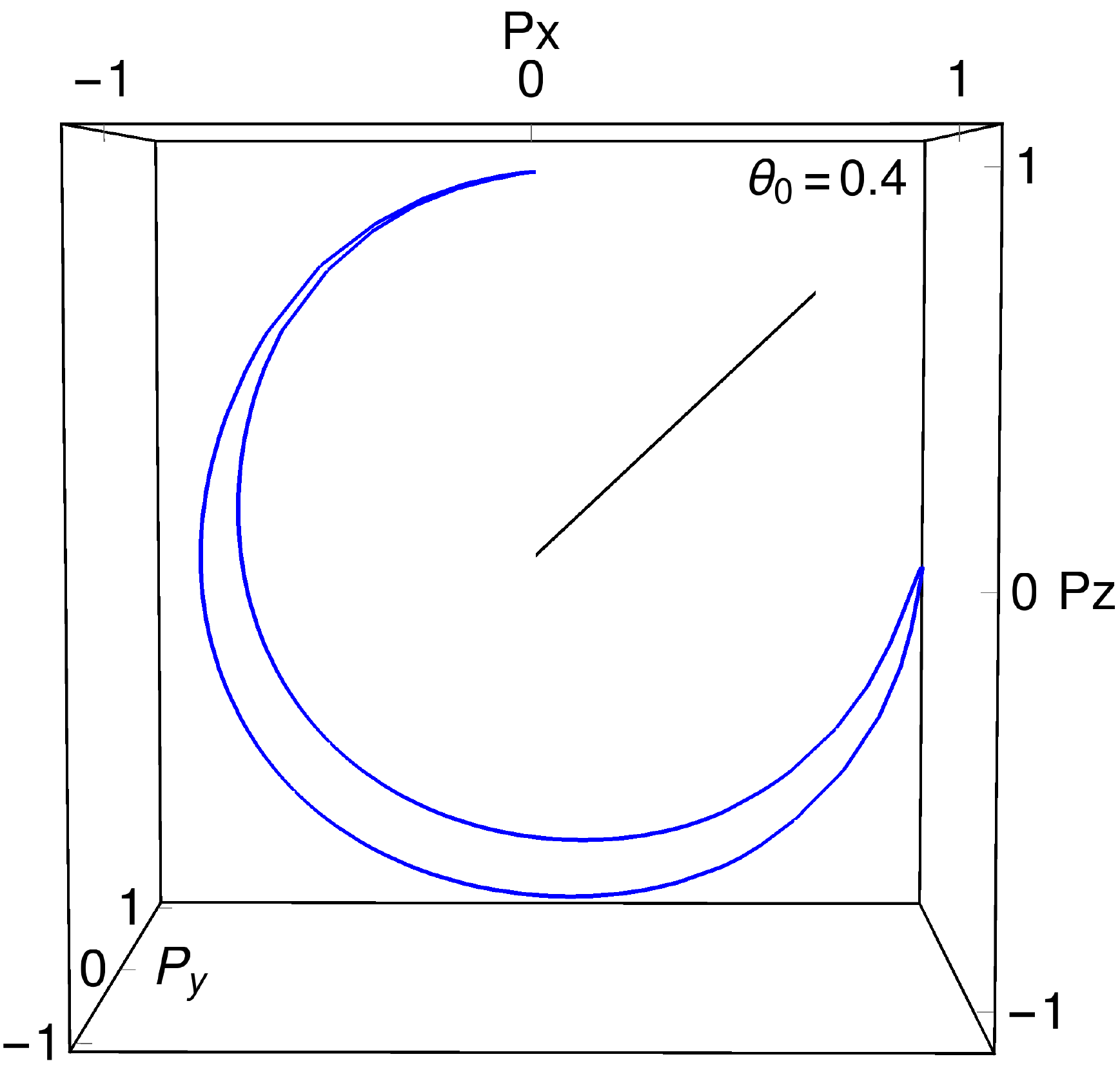}\includegraphics[scale=0.29]{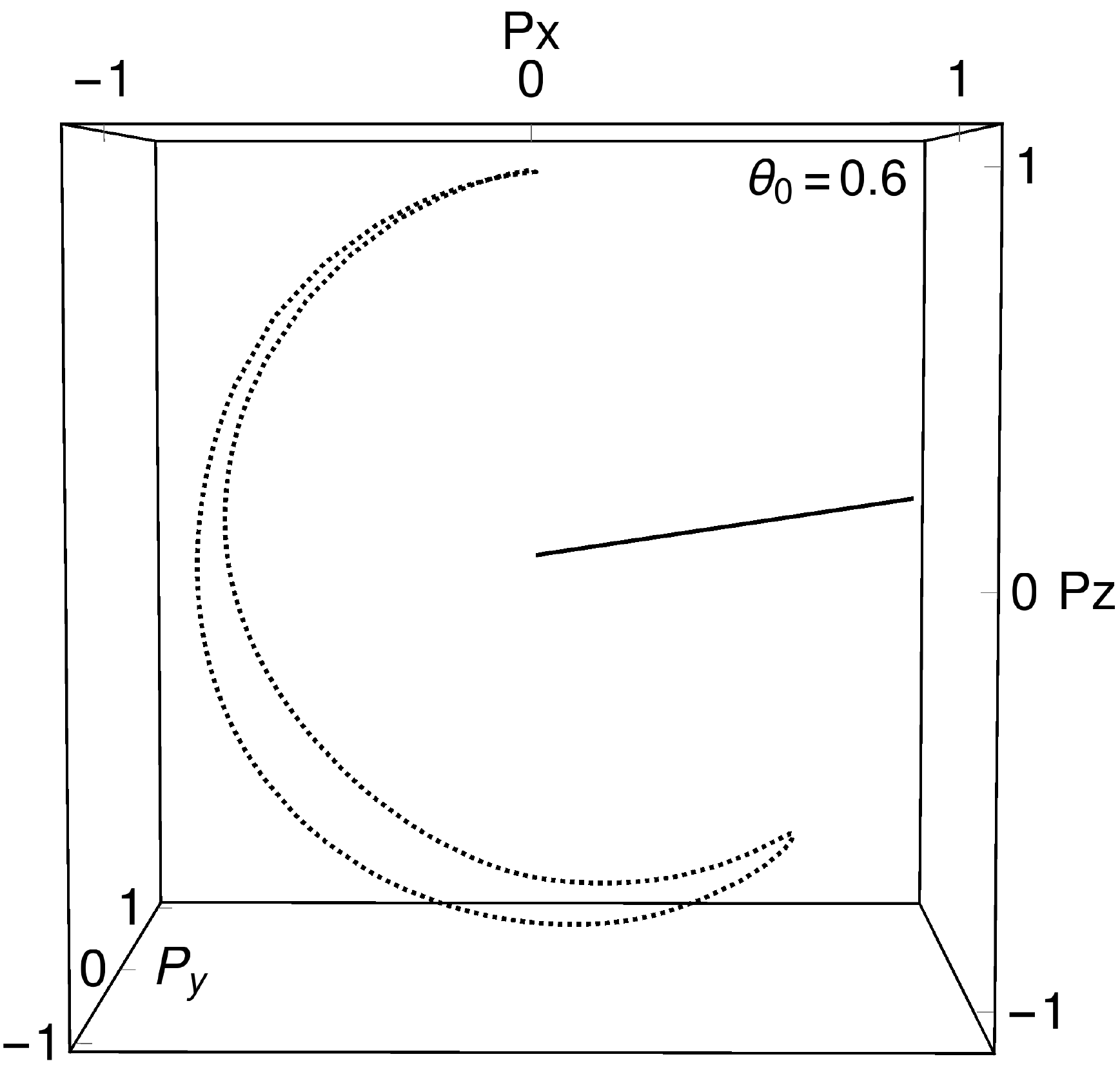}
\caption{Parametric plot of  $P_x(t),~P_y(t)$ and  $P_z(t)$ in IH for three different values of $\vartheta_0 = 0.01,0.4,0.6$ (corresponding to those in Fig.\,\ref{fig1a}). The pendular axis has also been shown. We take $g_3=1.5$ and $\mu = 10\,\w$.}
\label{fig1b}
\end{figure}
If the vacuum mixing angle $\vartheta_0$ is large (and $\lam=0$, so the effective mixing angle is not suppressed), the oscillations develop a doubly periodic pattern,
as shown in Fig.\,\ref{fig1a}. The oscillation wavelength remains the same, however a new ``double-dip'' structure is seen in the 
oscillation pattern.

This feature can be understood from the fact that when $\vartheta_0$ is large, the initial misalignment angle $2\vartheta_0$ is large.  
The pendular motion of ${\bf P}$, which initially starts from the $z$-axis, is then symmetric about this initial misalignment axis. However the motion of ${\bf P}$ is not exactly symmetric about the $z$-axis. This asymmetry becomes more prominent and visible when $\vartheta_0$ increases and this leads to the double-dip feature. If we realign our axis from which $\varphi$ is measured 
with the pendular axis, then these features would vanish.
In Eq.\,(\ref{eq:eom3}), this amounts to a shifting of $\varphi\rightarrow\widetilde{\varphi}=\varphi+2\vartheta_0$. 
 
 When the initial neutrino-antineutrino flux asymmetry is vanishing, and stays zero since $g_1=0$, we have $P_z(t)=\overline{P_z}(t)=S_z/2$, and Eq.\,(\ref{Qvec}) gives
 \begin{equation}\label{ddip}
  P_z=\frac{1}{2}\left(|{\bf Q}|\cos\varphi+\frac{\w}{\widetilde{\mu}}\cos2\vartheta_0\right)\,.
 \end{equation}
The maxima of $P_z$ occur at $\varphi_{\rm max,1}=\sin^{-1}\left[-\omega\sin2\vartheta_0\,/\left(\widetilde{\mu}|{\bf Q}|\right)\right]$ and $\varphi_{\rm max,2}=2\pi-4\vartheta_0+\varphi_{\rm max,1}$\,, while both the minima are at $\varphi_{\min}=\pi$. Equation\,(\ref{ddip}) then explains the double-dip feature. The heights of the two maxima are different, the larger maxima corresponding to $\varphi_{\rm max,1}$ and the smaller maxima corresponding at $\varphi_{\rm max,2}$.

 A clearer idea may be obtained if we study the 
motion of ${\bf P}$ in the 3-dimensional $\left[P_x(t),P_y(t),P_z(t)\right]$ space as shown in Fig.\,\ref{fig1b} .
When $\vartheta_0$ is small, the motion is almost symmetric about the $z$- axis, tracing an almost complete circle in the $P_x(t)-P_z(t)$ plane. It is interesting to note that while the pendulum comes back to its initial position, it does not retrace its path.
For larger $\vartheta_0$, the axis tilts significantly, and ${\bf P}$ traces a trajectory symmetric about this new axis. The double-dip features are therefore a result of an initial large misalignment.

\subsection{Neutrino-neutrino potential in a supernova} 
\label{varMat}
Inside a core-collapse supernova, the neutrinos will experience a time-varying potential
as they travel outwards from the neutrinosphere. Moreover, the initial neutrino flux is typically more than the antineutrino flux. To take into account these features, we consider the following spherically symmetric potential \cite{Dasgupta:2009mg},
\begin{equation}\label{mu}
\mu=7.5\times10^5~{\rm km^{-1}}~\left(\dfrac{r_0}{r}\right)^4\,, \qquad r>r_0\,, \\
\end{equation}
where $r_0=10~\text{km}$ is taken to be the radius of the neutrinosphere. We choose 
$\w=0.3~{\rm km}^{-1}$ corresponding to the  atmospheric mass squared difference and  $E\simeq 20$ MeV, and a neutrino-antineutrino flux asymmetry of $20\%$. We take $\lam=0$ and study the effects of FP-NSSI and FV-NSSI through the rescaled couplings $g_3$ and $g_1$ (see Eq.\,(\ref{rescale})), respectively.

\begin{figure}[!t]
\begin{center}
  \includegraphics[width=0.48\textwidth]{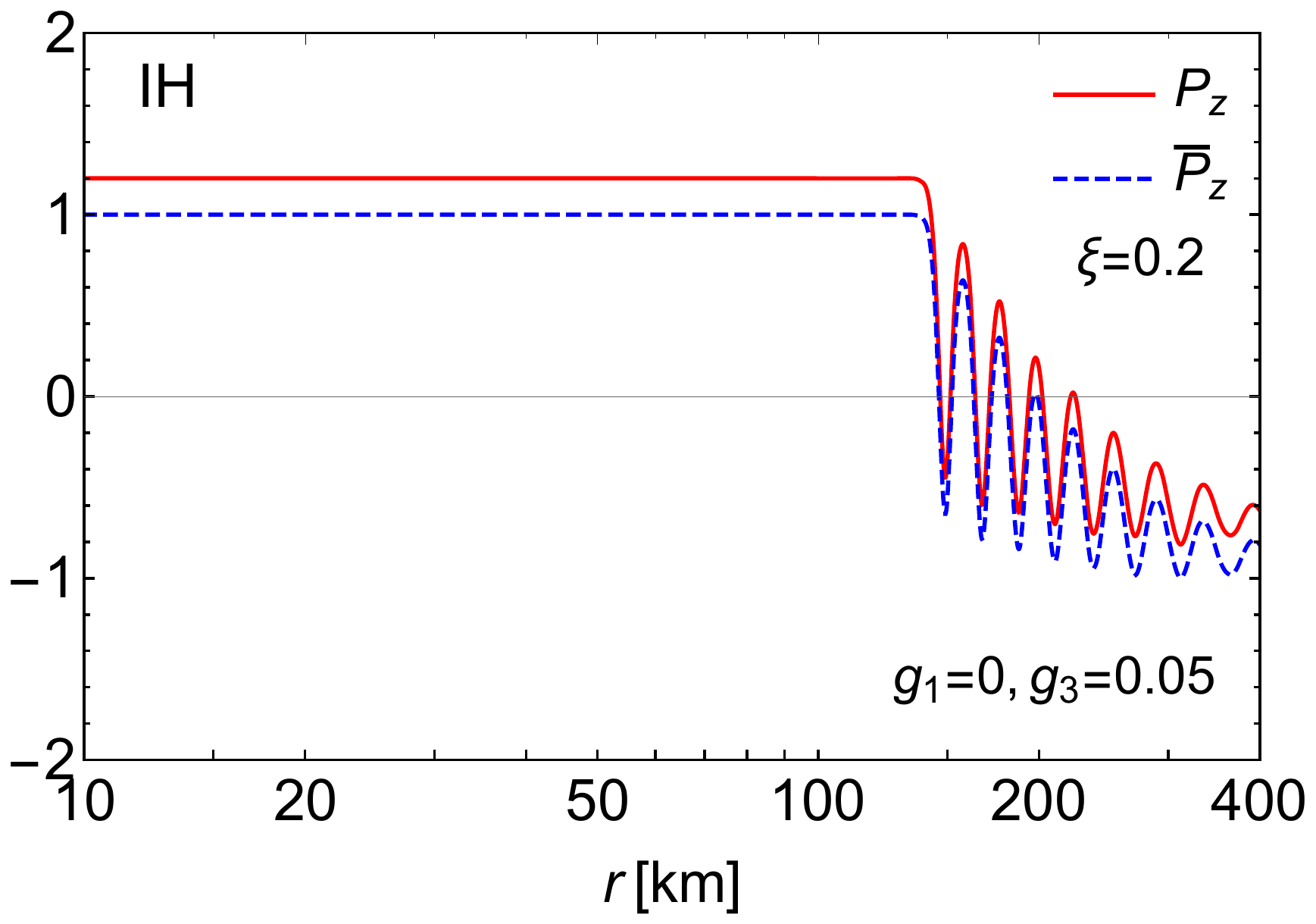}~~~~~~\includegraphics[width=0.466\textwidth]{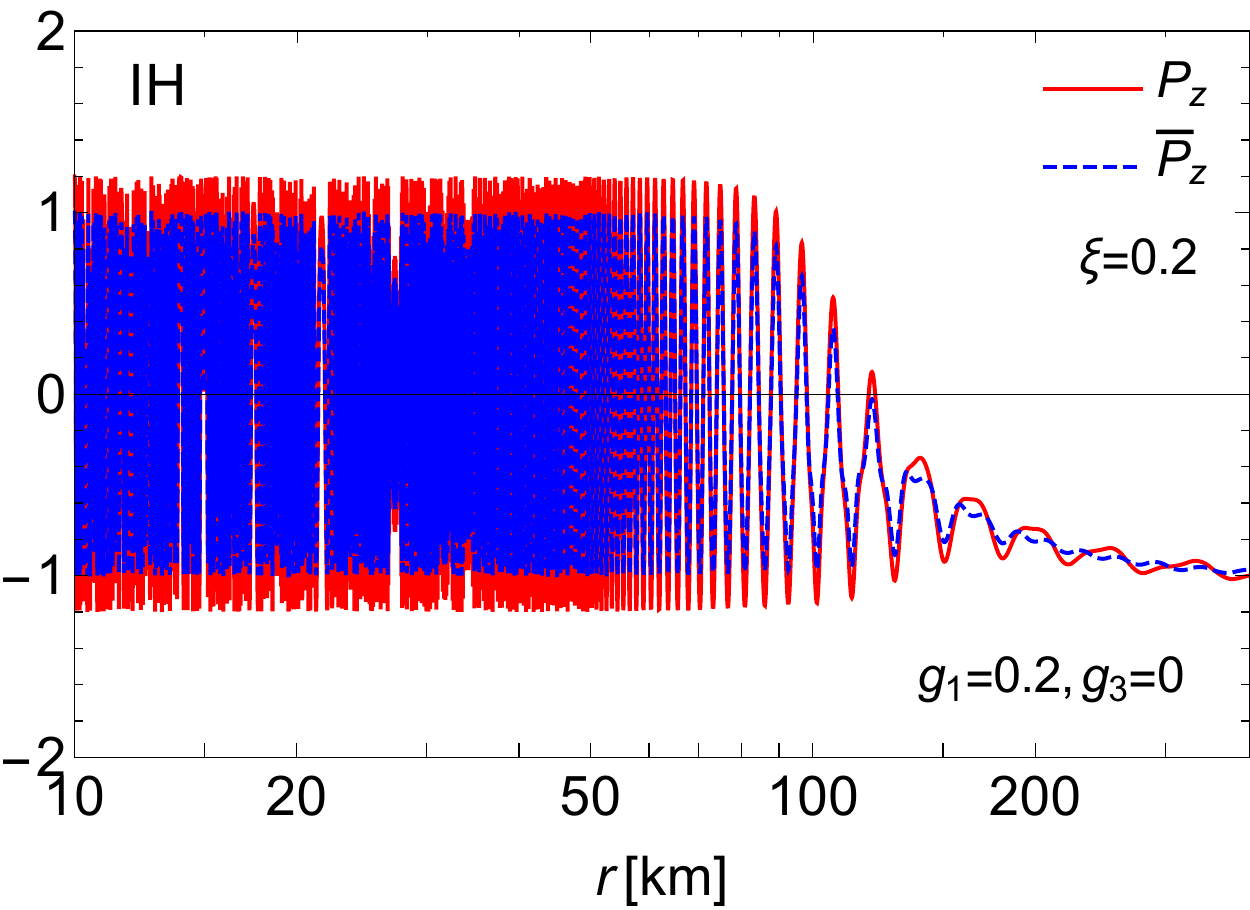}
\caption{Time evolution of $P_z$ and $\overline{P}_z$ in IH with decreasing neutrino potential as in Eq.\,(\ref{mu}). The asymmetry is chosen to be $\xi=0.2$ and $\w=0.3~{\rm km}^{-1}$. Left: $g_1\,=\,0,\,g_3\,=\,0.05$. Right:  $g_1=0.2,\, g_3=0$. }
\label{fig3}
\end{center}
\end{figure}

 We show the evolution of  $P_z$ and $\overline{P}_z$ in
Fig.\,\ref{fig3}. The polarization vectors have been normalized such that $|\overline{{\bf P}}|=1$. The following observations may be made from the figure.
\begin{itemize}
 \item FP-NSSI scenario: When only $g_3$ is non-zero , the initial flux asymmetry is conserved, a result of ${\bf B}\cdot{\bf D}$ conservation.  The initial flat values of $P_z$ and $\overline{P}_z$ denotes synchronized oscillations. It is important to note that the presence of $g_3$ leads to an extension of the time of onset of bipolar oscillations as explained in Section \ref{sma}. The bipolar oscillations, which would have started at $r\simeq100~{\rm km}$ at $g_3=0$,
start at $r \simeq 130~{\rm km}$ now for $g_3=0.05$. Beyond that, almost complete flavor conversion takes place, while conserving the total flavor lepton number, hence retaining the flux asymmetry.

 \item FV-NSSI scenario: When only $g_1$ is non-zero, the flavor lepton number is not conserved. 
Rapid oscillations are observed to take place even at very low $r$ values, due to the ``transverse'' NSSI term proportional to $\xi\mu g_1\,\hat{\bf x}$.
With increasing $r$, the value of $\mu$ decreases and so does the frequency of oscillation. During this evolution, the value of the flux asymmetry $\xi$ keeps on changing. It is finally frozen at large $r$ when $\xi\mu g_1\rightarrow 0$. Even in this scenario, almost complete flavor conversion may take place.
\end{itemize}
Thus, we find that even in the simple single-angle approximation, the introduction of NSSI gives rise interesting results in the time evolution
of system of neutrinos and antineutrinos of a single energy. In the next section, we will focus on a spectrum of 
neutrinos and antineutrinos, and study the effects of FP-NSSI and FV-NSSI on it.

%%%%%%%%%%%%%%%%%%%%%%%%%%%%%%%%%%%%%%%%%%%%%%%%%%%%%%%%%%%%%%%%%%%%%%%%%%%%%%%%%%%%%%%%%%%%%%%%%%%%%%%%%%%%%%%%%%%%%%%%%%%%%%%%%%%%%%%%%%%%%%%%%%%%%%%%%%%%%%%%%%%%%%%%%%%%%%%%%%%%%%%%

\section{Effects of NSSI on spectral swaps }
\label{sec:3}
\begin{figure}[!t]
\begin{center}
\includegraphics[scale=0.42]{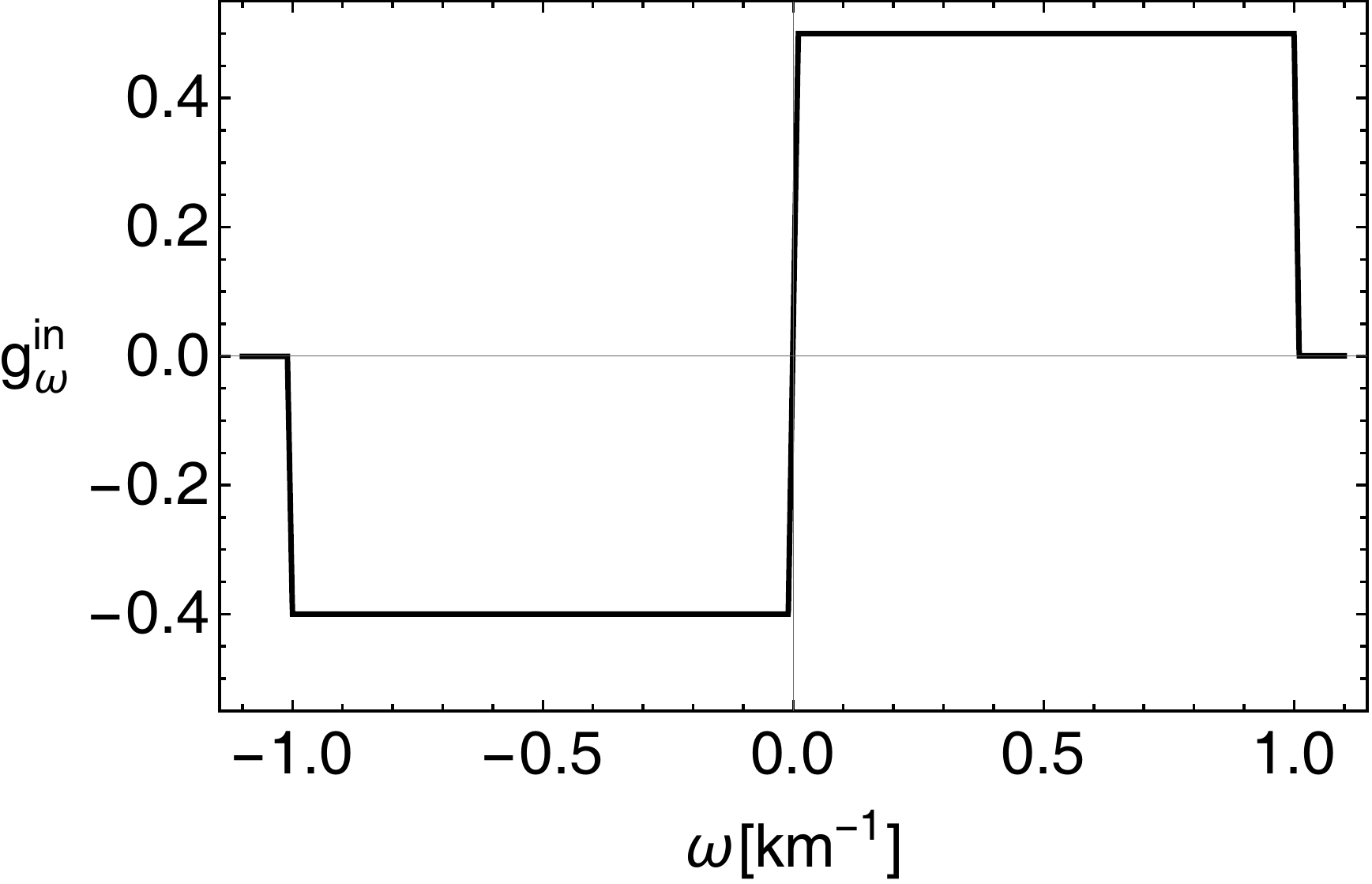}\quad\includegraphics[scale=0.42]{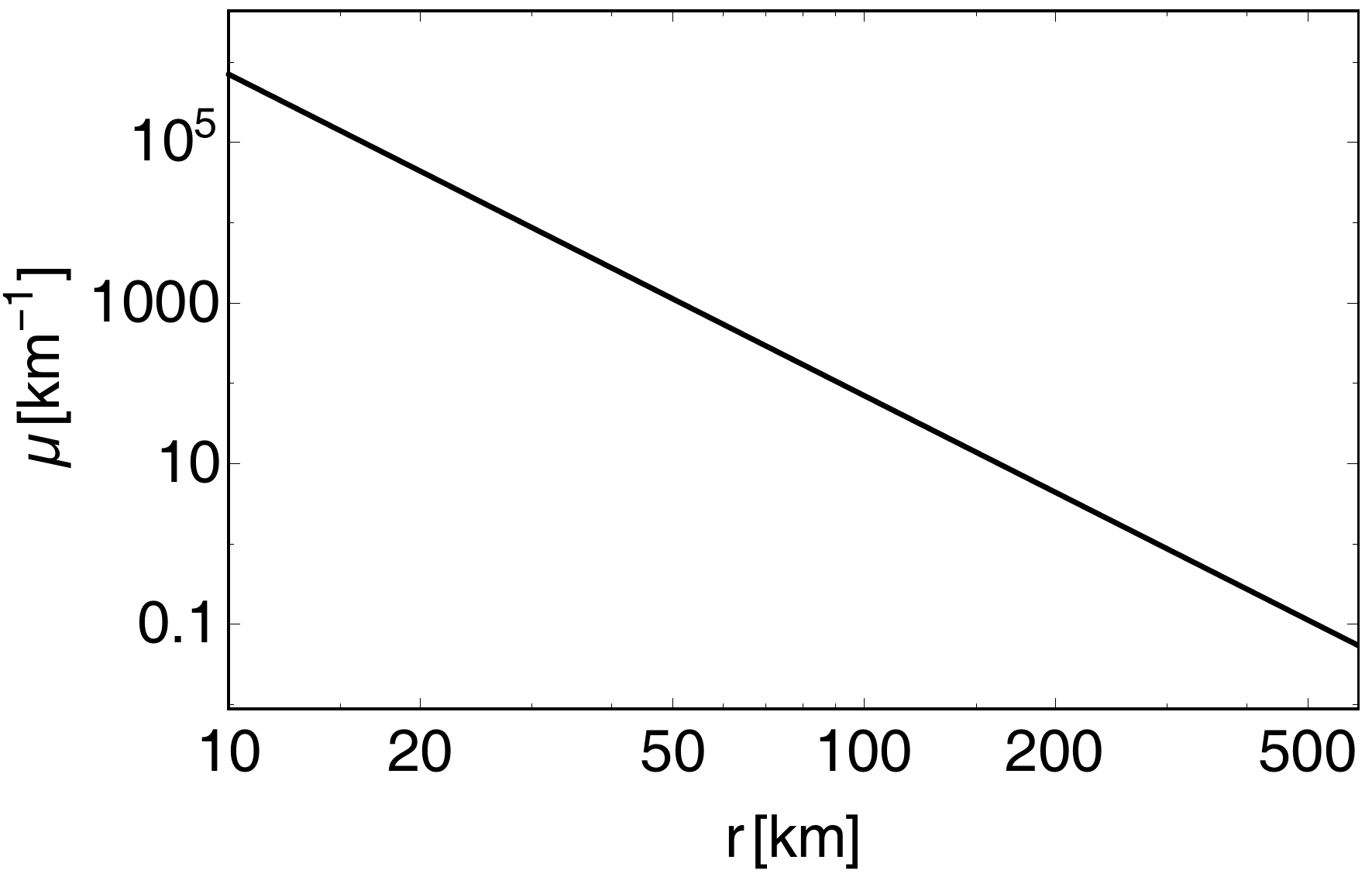}
\caption{Initial spectrum  and the background neutrino-neutrino potential for the analysis in Section~\ref{sec:3}.
Left: Initial spectrum $g_\w^{\rm in}$. Right: The neutrino-neutrino potential $\mu(r)$. }
\label{fig4a}
\end{center}
\end{figure}

Till now, we have studied the effects of NSSI on a single energy mode. In
this section, we explore its effects on a toy spectrum, over a range of $\w$ values. This will help us get a clear understanding
of how NSSI can affect the spectral swaps.
%Then absence of NSSI may be considered to be a single point $(0,0)$ in the $g_1-g_3$ plane.
In this section, we explore the NSSI effects in the $g_1-g_3$ plane and 
illustrate our results with a box spectrum $g_\w^{\rm in}$ (see Eq.\,(\ref{spectrum})). The spectrum, shown in Fig.\,\ref{fig4a} (left panel), corresponds to a flat $\w$ spectrum of neutrinos and antineutrinos, confined to $0\leq|\w|\leq 1$,
with a $\nu-\bar{\nu}$ flux asymmetry of $10\%$. The neutrino flavors evolve while propagating through a medium with the neutrino-neutrino potential $\mu$ described by Eq.\,(\ref{mu}) and shown in Fig.\,\ref{fig4a} (right panel).

\subsection{FP-NSSI scenario: pinching of spectral swaps } 
\label{FP}
\begin{figure}[!t]
  \includegraphics[scale=0.444]{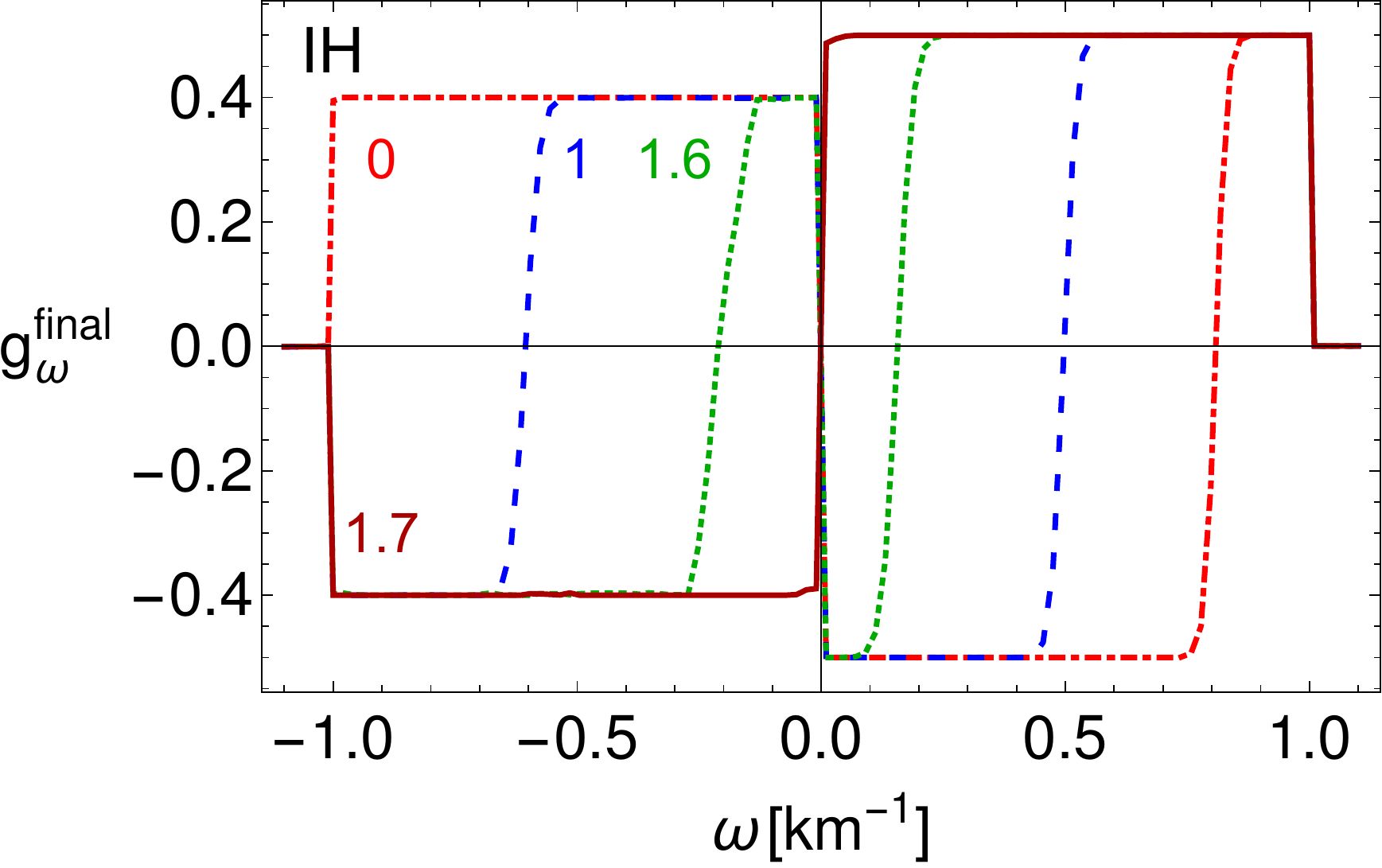}~  \includegraphics[scale=0.43]{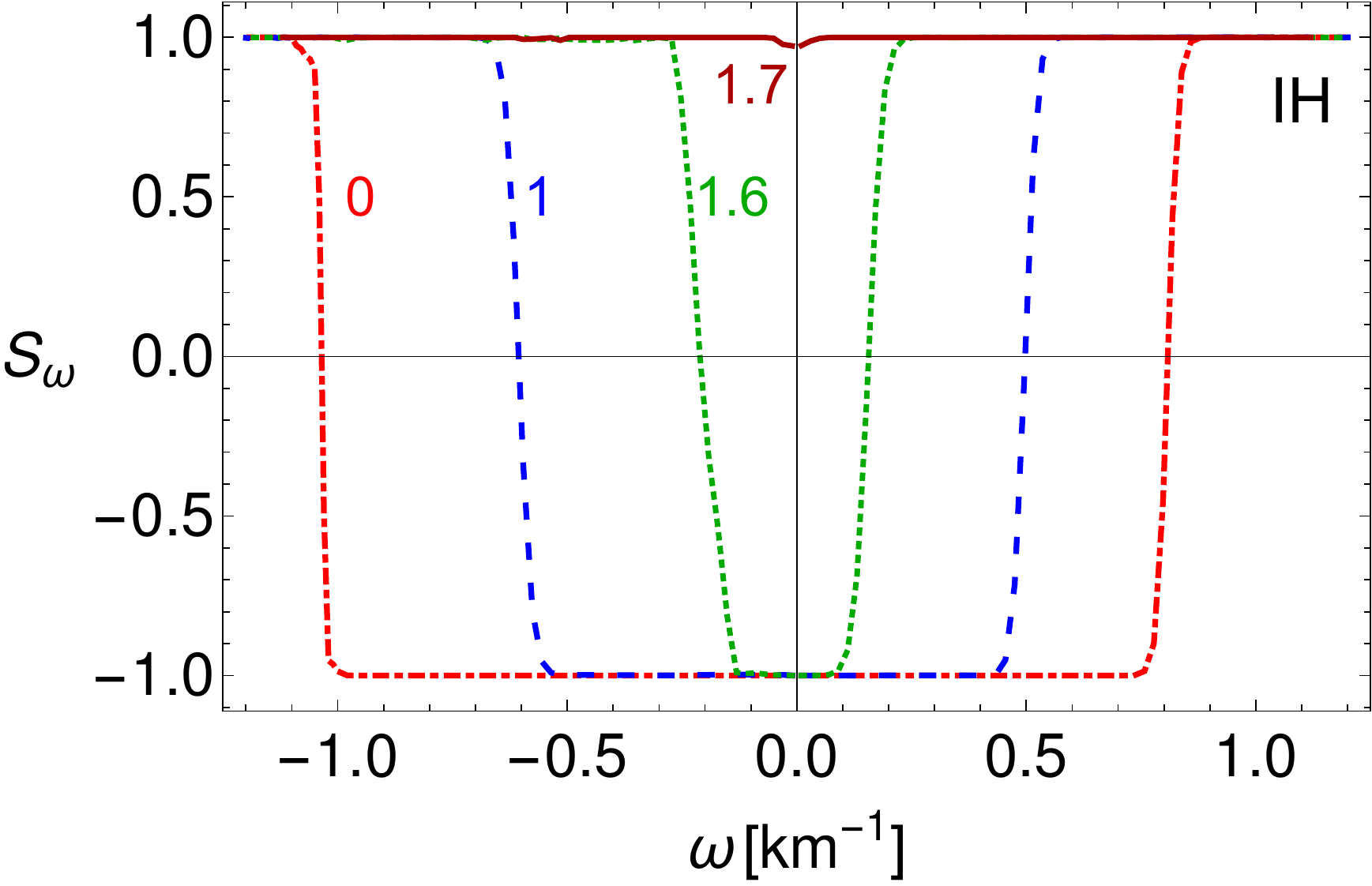}
\caption{The final spectrum $g_\w^{\rm final}$ (left panel) and the swap factor $S_\w$ (right panel) for $g_1=0$ and $g_3=0,~1,~1.6,~1.7$ .}
\label{fig4}
\end{figure}

When $g_1=0,~g_3\neq0$, the NSSI terms play the same role in the EoMs as the matter term, as we have seen in Sections~\ref{sma} and \ref{varMat}. 
It is already known that presence of a matter term tends to suppress collective oscillations~\cite{Chakraborty:2011nf}. 
So we expect that increasing values of  $g_3$ would lead to suppression of spectral swaps. In Fig.\,\ref{fig4}, we
show the final spectrum $g_\w^{\rm final}$ (left panel) and the swap factor $S_\w$ (right panel) for different values of $g_3$, in IH.

It is observed that the swaps develop around 
the positive crossing at $\w=0$, with their relative extent on both sides of the crossing 
being fixed by approximate conservation of ${\bf B}\cdot{\bf D}$. The width of the swap is the greatest in the absence of $g_3$ (SM scenario). With increasing $g_3$, the width of the swap is observed to decrease, i.e., the swaps get pinched. As $g_3$ approaches $2$, the height of the swap also decreases till the swap finally vanishes when $g_3\simeq 1.7$. Note that the value of $g_3$ at which the swap vanishes depends on initial spectrum, however our arguments in Section~\ref{sma} have already indicated that for $g_3>2$ in IH, there would not be any collective oscillations. 

It is important to mention that in this case, the conservation of ${\bf B}\cdot{\bf D}$ is approximately valid in the limit of small mixing angle $\vartheta_0$. From Eq.\,(\ref{eq:eom2a}), we find
\begin{equation}\label{LNVg3}
 \frac{d}{dt}{\bf B}\cdot{\bf D}=\widetilde{\lambda}\left[{\bf B}\,{\boldsymbol g}\,{\bf D}\right]\,,
\end{equation}
where $[\cdots]$ represents the scalar triple product (box product).
If $\vartheta_0\simeq 0$, the vectors ${\bf B}$ and ${\boldsymbol g}=g_3\,\hat{\bf z}$ are nearly parallel, the box product vanishes, and the flavor lepton number conservation is valid.

The effect of the FP-NSSI coupling is thus to
reduce both the height and width of the spectral swaps.

\subsection{FV-NSSI scenario: flavor lepton number violation}
\begin{figure}[!t]
\begin{center}
  \includegraphics[scale=0.43]{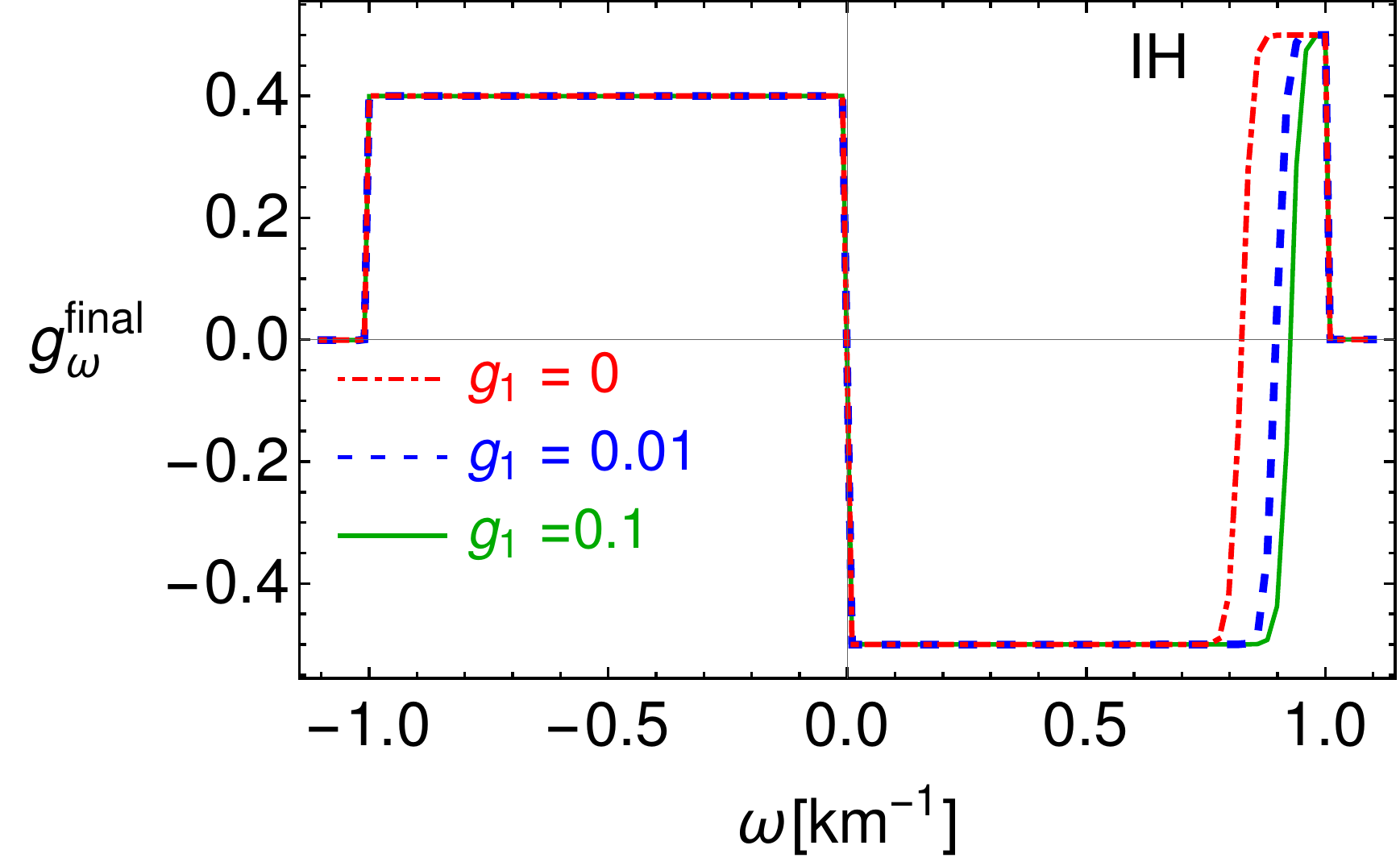}~  \includegraphics[scale=0.435]{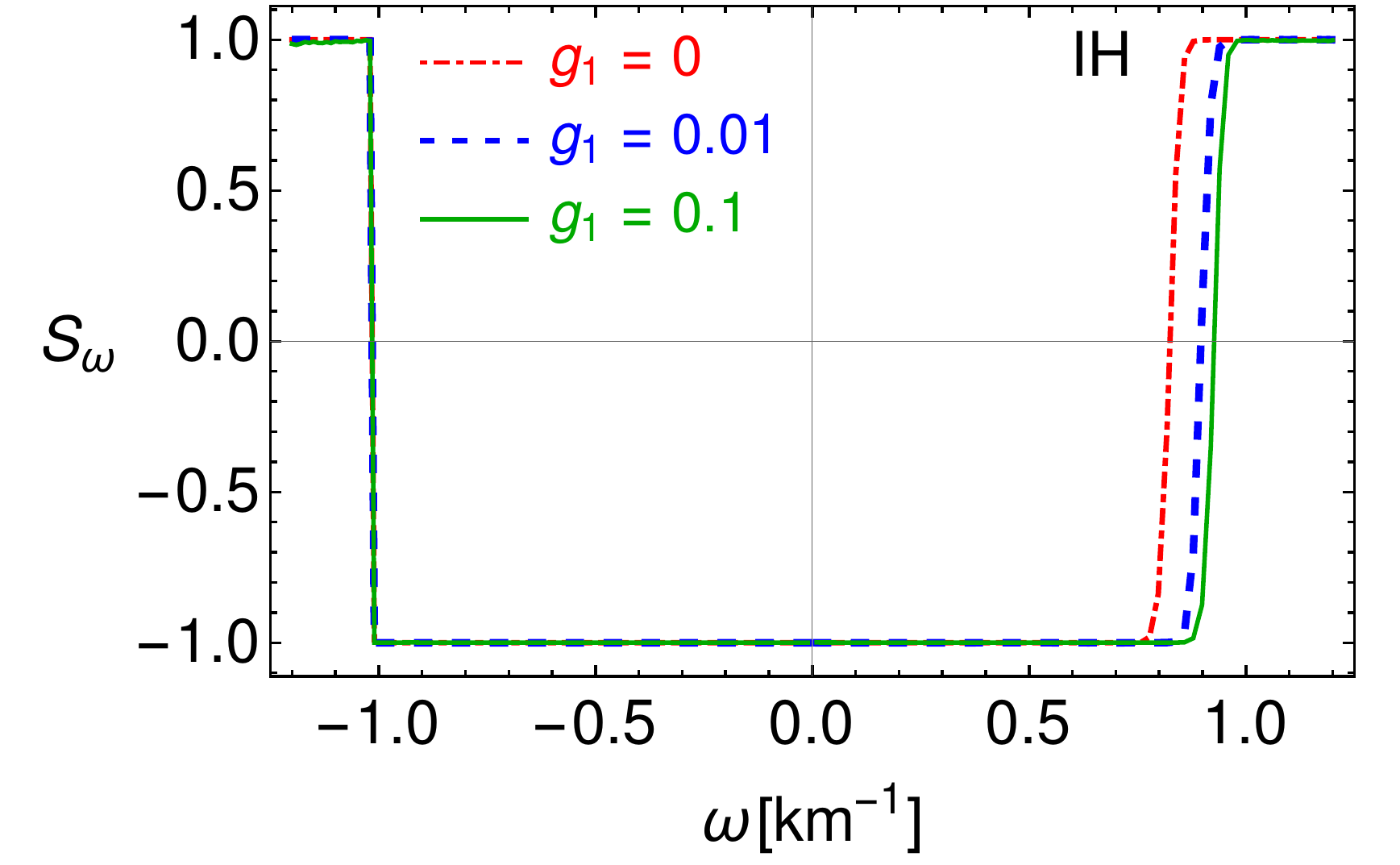}
\caption{The final spectrum $g_\w^{\rm final}$(left panel) and the final swap factor $S_\w$ (right panel) for different values of $g_1=0,~0.01,~0.1$ and $g_3=0$.}
\label{fig5}
\end{center}
\end{figure}

The presence of a non-zero $g_1$ provides a more interesting scenario to study. The term $\xi\mu g_1\,\hat{\bf x}$ essentially acts like an oscillation term
between the two neutrino flavors. It causes flavor lepton number violation
and can give rise to flavor conversions even with $\vartheta_0=0$ \cite{Blennow:2008er}. Our results in Section \ref{sec:2}, where we plot the variation of $P_z,\overline{P}_z$ in Fig.\,\ref{fig3} (right panel), also illustrate the non-conservation of flavor lepton number.
Indeed with ${\boldsymbol g}=g_1\,\hat{\bf x}$,  Eq.\,(\ref{LNVg3}) gives
\begin{equation}\label{LNV2}
 \frac{d}{dt}{\bf B}\cdot{\bf D}=-\widetilde{\lambda}g_1 D_y \cos2\vartheta_0\, ,
\end{equation}
which is non-zero even for $\vartheta_0=0$. This non-conservation of ${\bf B}\cdot{\bf D}$ may have important observations. We illustrate the effect of $g_1$ on the same spectra as in Section~\ref{FP} in Fig.\,\ref{fig5}.

The figure shows that the effect of $g_1$ is felt at very low values. With our particular spectrum,
increasing $g_1$ increases the width for $\w>0$, while keeping the swap for $\w<0$ unchanged. Clearly this implies that ${\bf B}\cdot{\bf D}=\int d\w g_\w$ is not conserved.
The width of the swap on
both sides of the split is now governed by the variation in Eq.\,(\ref{LNV2}).

\begin{figure}[!h]
\centering
  \includegraphics[width=0.5\textwidth, height=0.305\textwidth]{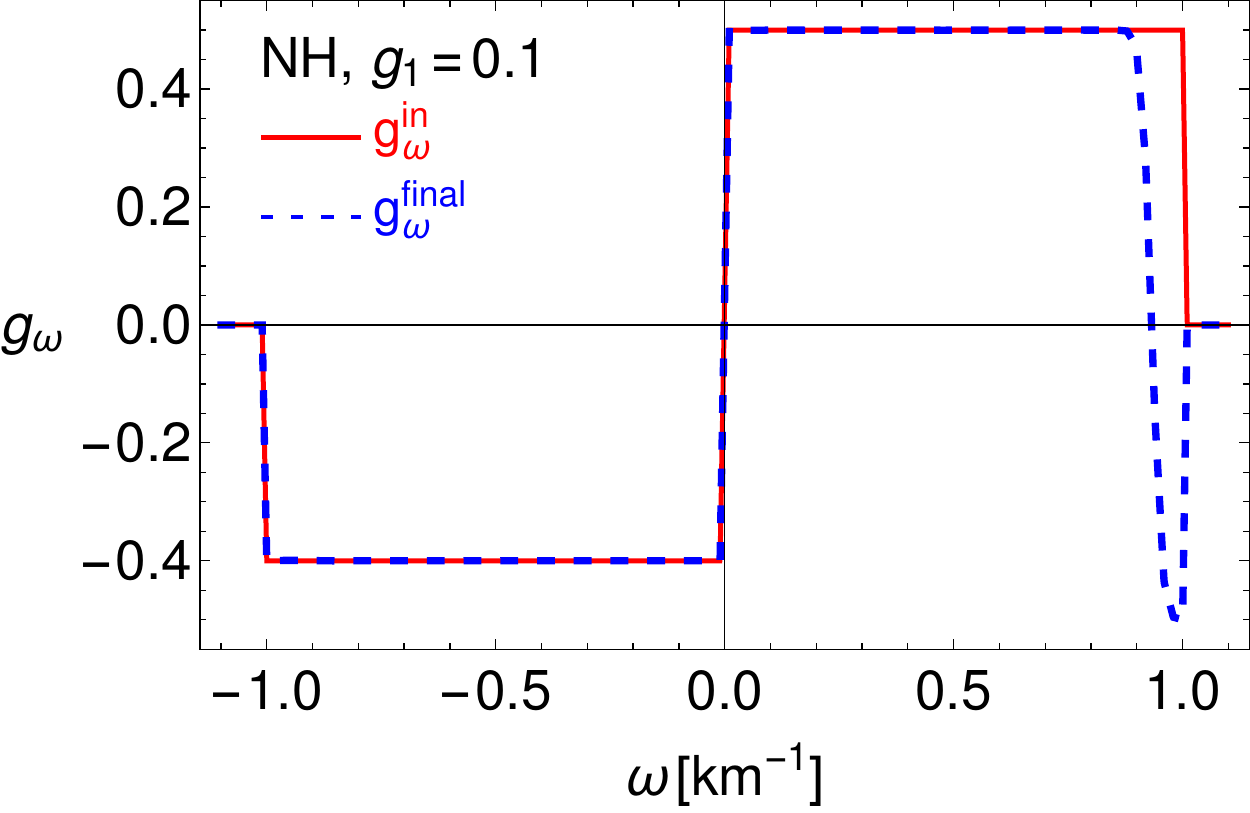}~\includegraphics[width=0.5\textwidth, height=0.3068\textwidth]{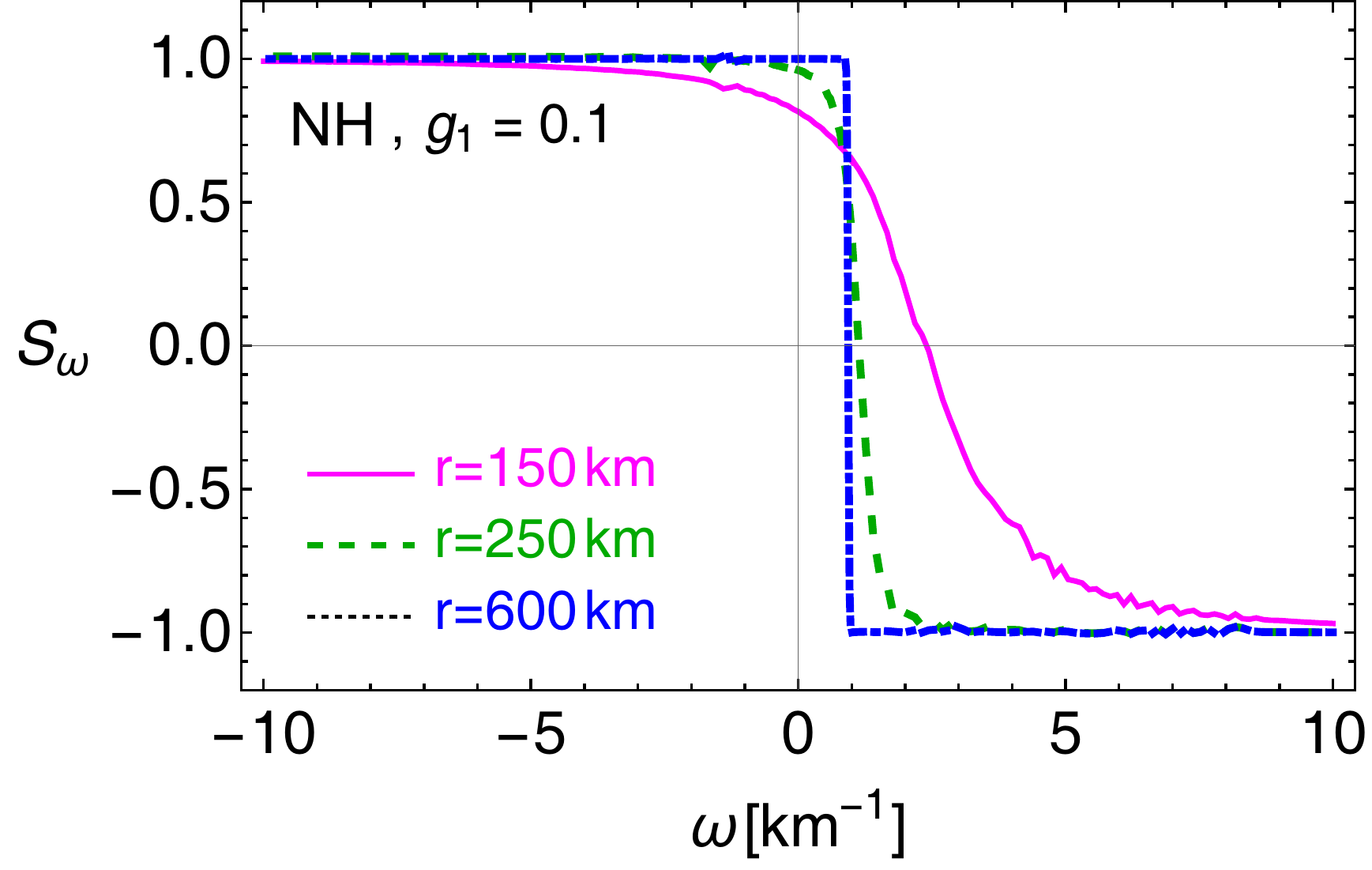}
\caption{Left panel: The initial (red) and final (blue) spectrum in NH. Right panel: The development of the swap factor $S_\w$ at different distances from the neutrinosphere for $g_1=0.1$.}
\label{fig6}
\end{figure}

An important observation may be made at this stage.
When ${\bf B}\cdot{\bf D}$ is conserved, $ \int d\w g_\w$ over the swapped region should vanish. This implies that the swapped region must contain at least one zero crossing in $g_\w$, by the intermediate value theorem.
This is the essential reason for swaps to develop around a crossing. Therefore in the SM as well as in FP-NSSI, collective oscillations start taking place only when there is a $g_\w$ crossing, which is the common wisdom.
However, this notion breaks down when we have FV-NSSI.  Now ${\bf B}\cdot{\bf D}$ is not conserved anymore 
and, depending on the evolution in Eq.\,(\ref{LNV2}), swaps may occur away from the zero crossing of $g_\w$. 

We illustrate an interesting consequences of this observation in Fig.\,\ref{fig6}, where we take the same initial spectrum $g_\w^{\rm in}$ as in Fig.\,\ref{fig4a} (left panel), however consider the NH case. In the SM, we know that a swap cannot develop in such a case since the crossing at $\w=0$ is positive. Even the discontinuity at $\w=1$ is not a real negative crossing and  cannot lead to a swap as long as ${\bf B}\cdot{\bf D}$ is conserved.
However, as the Fig.\,\ref{fig6} (right panel) shows, in presence of $g_1$,
a swap starts developing for $\w\simeq 1$, resulting in collective oscillations in that region.
To understand the evolution of this swap further, we have added a $g_\w=10^{-5}$ (for $1\leq\w\leq10$) and observed the development of the swap at different distances from the neutrinosphere. As the right panel of Fig.\,\ref{fig6} shows, the swap clearly starts developing beyond the discontinuity at $\w=1$ where the spectrum virtually vanishes, a feature not seen in the case of the SM. Interestingly, the swap factor remains at $S_\w=-1$  and does not change to $S_\w=1$ even for large enough $\w$.

Thus, introduction of an off-diagonal NSSI leads to unexpected swaps in the neutrino spectrum.
This flavor  number non-conservation can also lead to new spectral splits even in
the neutronization burst epoch, as we will see in Section \ref{sec:4}.

%%%%%%%%%%%%%%%%%%%%%%%%%%%%%%%%%%%%%%%%%%%%%%%%%%%%%%%%%%%%%%%%%%%%%%%%%%%%%%%%%%%%%%%%%%%%%%%%%%%%%%%%%%%%%%%%%%%%%%%%%%%%%%%%%%%%%%%%%%%%%%%%%%%%%%%%%%%%%%%%%%%%%%%%%%%%%%%%%%%%%%%%

\section{Collective effects during neutronization burst}
\label{sec:4}

In the SM, collective effects cause pairwise conversions $\nu_e\leftrightarrow\nu_\alpha$ and $\bar{\nu}_e\leftrightarrow\bar{\nu}_\alpha$ due to flavor lepton number conservation.
During the neutronization burst, only $\nu_e$s are present. As a result, there is no zero-crossing in the spectrum $g_\w$ and hence
bipolar oscillations do not occur. However, as we have shown in Section \ref{sec:3}, the presence of FV-NSSI can provide the necessary
seed for collective oscillations to develop. As a result, spectral splits may be observed 
even during the neutronization burst. Note that during this epoch, none of the other collective effects, including the fast flavor conversions, can give rise to this phenomenon.

\begin{figure}[!t]
\centering
  \includegraphics[width=0.4945\textwidth]{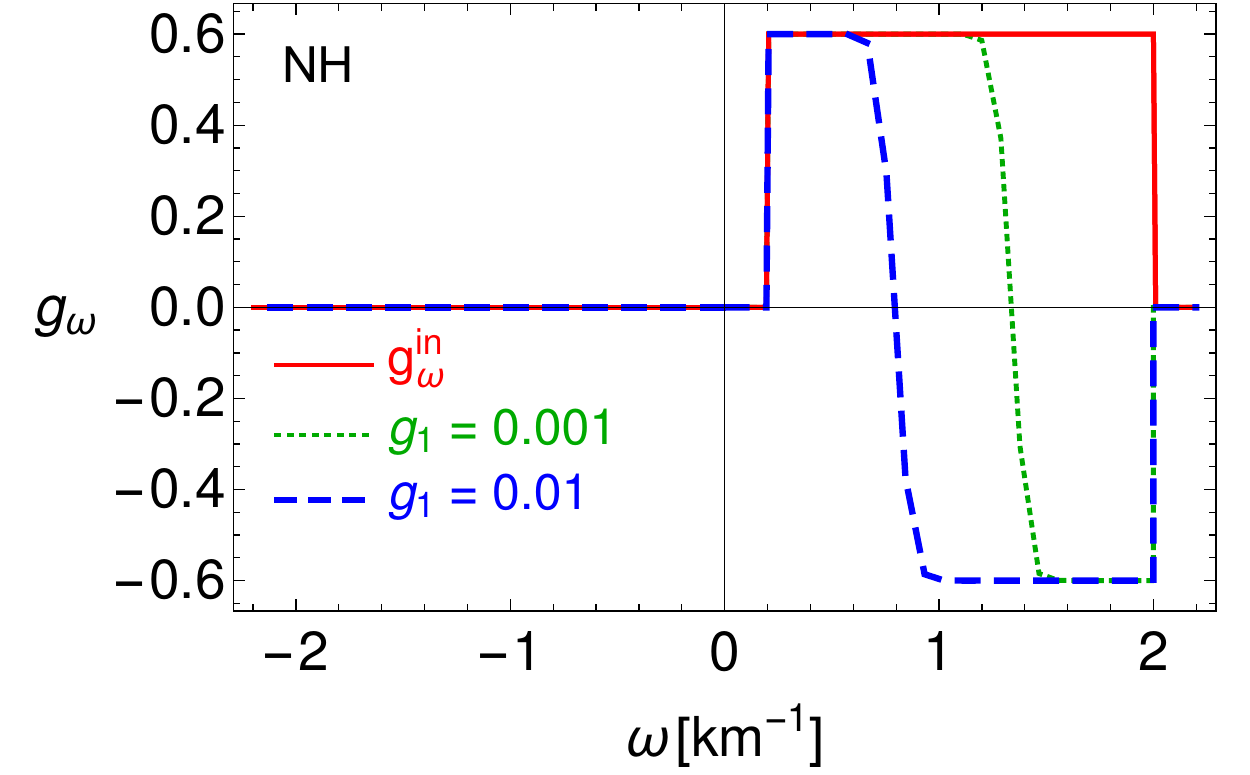}
  \includegraphics[width=0.498\textwidth, height=0.309\textwidth]{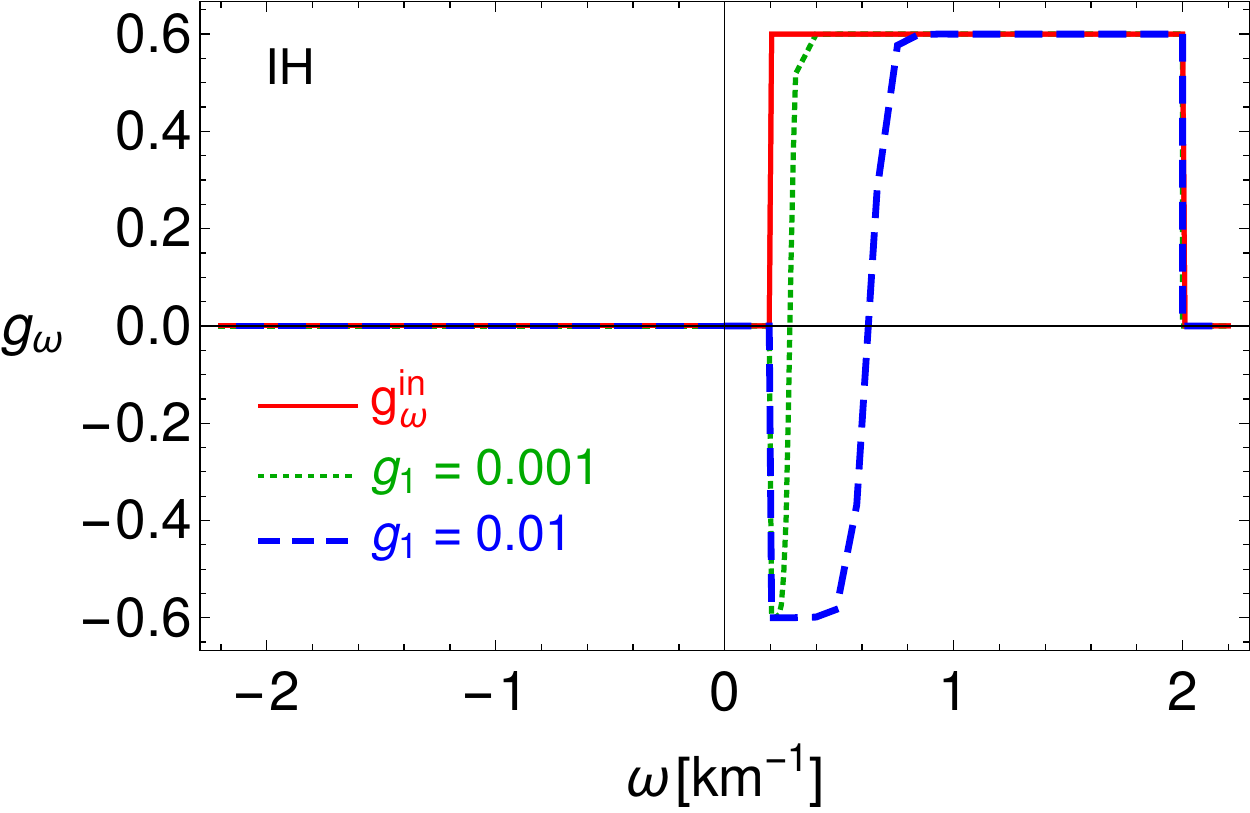}
   \includegraphics[width=0.478\textwidth]{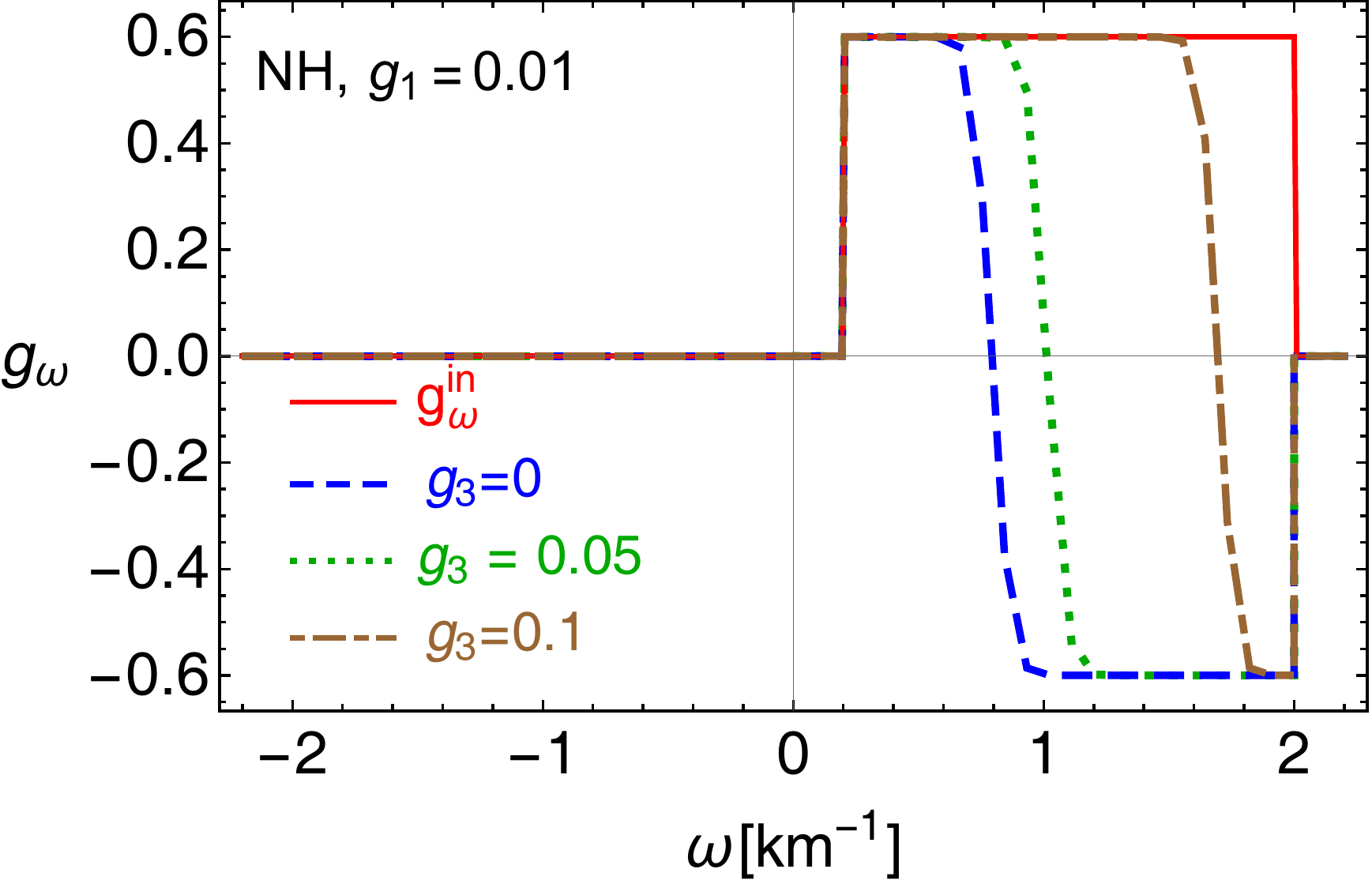}~~~
  \includegraphics[width=0.482\textwidth, height=0.31\textwidth]{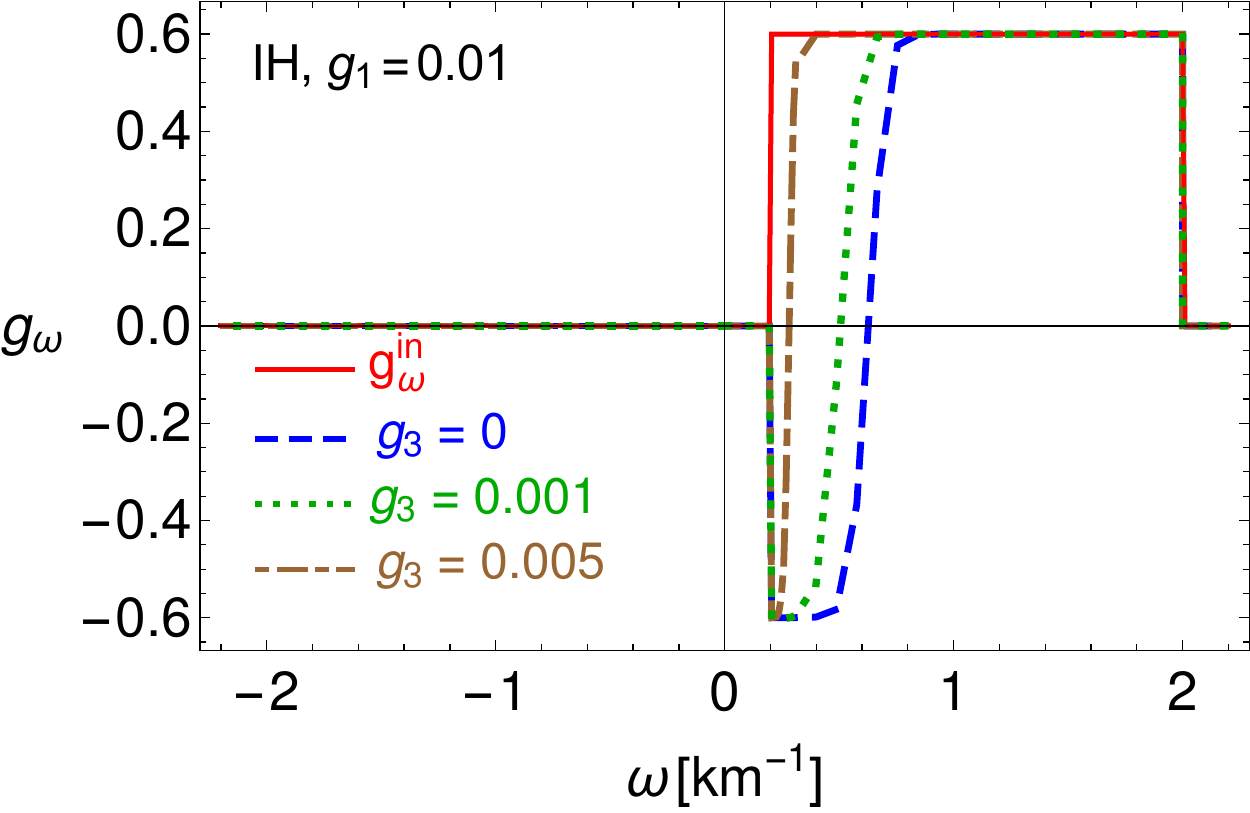}
\caption{Effects of collective oscillations due to NSSI on a pure $\nu_e$ spectrum restricted to $0.2\leq\w\leq2$, shown in solid line. 
Left Panels: NH, Right Panels: IH. Top Panels: The final spectra  with  $g_3=0$ (solid) and $g_1=0.001$ (dotted), and  $g_1=0.01$ (dashed). 
Bottom Panels: The pinching of final spectra for $g_1=0.01$  with $g_3=0$ (dashed), $g_3=0.05$ (dotted) and  $g_3=0.1$ (dotdashed) in NH (left). Similar plot in IH (right) for $g_1=0.01$ with $g_3=0$ (dashed), $g_3=0.001$ (dotted) and $g_3=0.005$ (dotdashed).}
\label{fig8}
\end{figure}

In Fig.\,\ref{fig8}, we demonstrate
this novel phenomenon with an initial box spectrum consisting only of $\nu_e$s.
This spectrum is non-zero for $\w_{\rm min}\leq\w\leq \w_{\rm max}$, which represents a cut-off in the $\nu_e$ spectra at low and high energies. We observe that
\begin{itemize}
 \item In NH (left panels),
the presence of a non-zero $g_1$ leads to the development of a swap around $\w_{\rm max}$, which corresponds to
conversions of low energy $\nu_e$s to $\nu_\alpha$s. With increasing value of $g_1$, the swap becomes broader, thereby converting more of the $\nu_e$s to $\nu_\alpha$s. For the spectrum used here, this phenomenon is visible for a non-zero FV-NSSI as low as $10^{-3}$.

\item In IH (right panels),
the presence of a non-zero $g_1$ leads to the development of a swap around $\w_{\rm min}$, which corresponds to
conversions of high energy $\nu_e$s to $\nu_\alpha$s. With increasing value of $g_1$, the swap becomes broader, thereby converting more of the  $\nu_e$s to $\nu_\alpha$s. For the spectrum used here, this phenomenon is visible for a non-zero FV-NSSI as low as $10^{-3}$.

\item If a non-zero $g_3$ is also present, the swap is pinched, as shown in the lower panels.
 
\end{itemize}

\begin{figure}[!t]
 \includegraphics[width=0.48\textwidth]{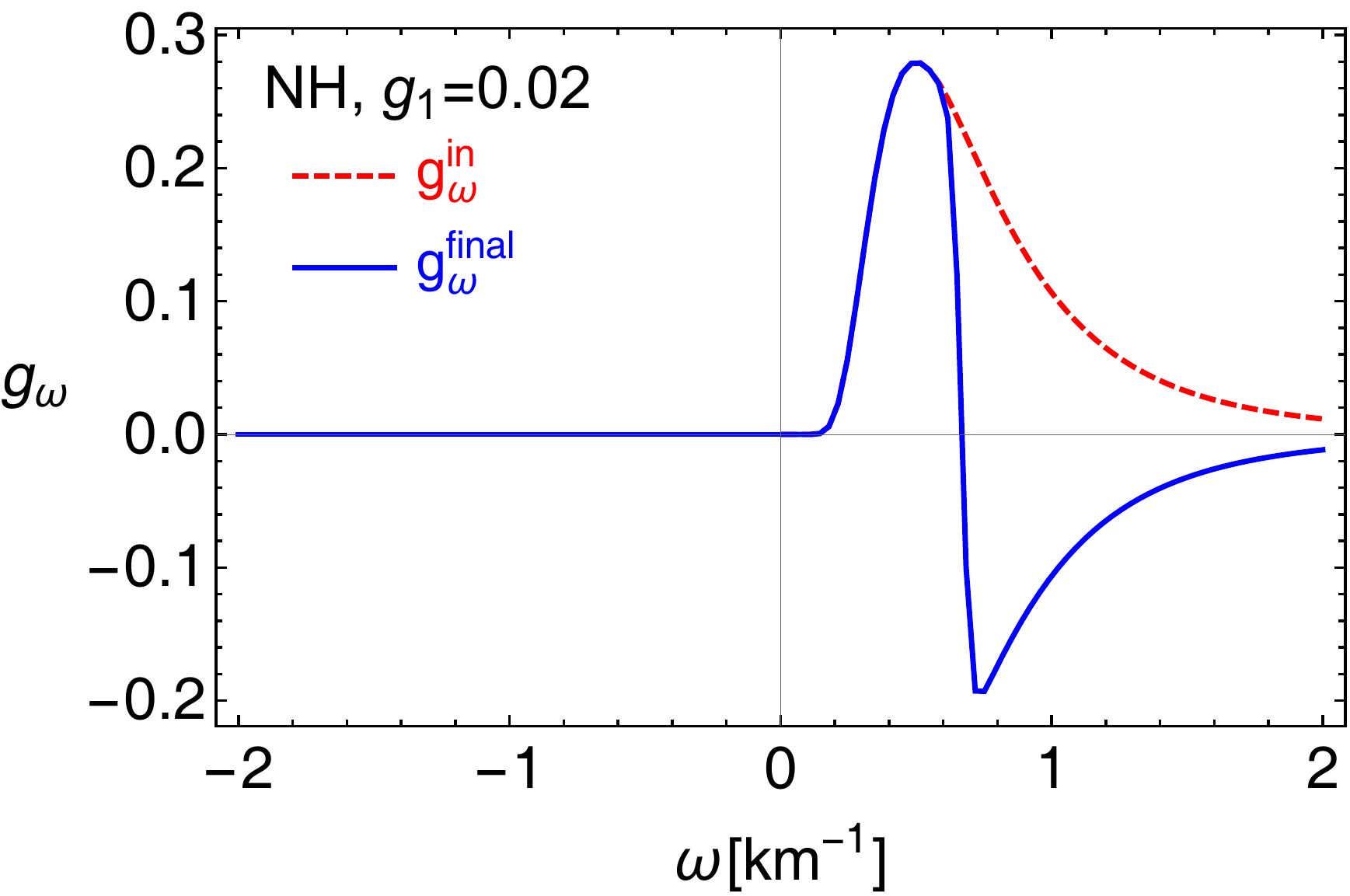}
 \includegraphics[width=0.48\textwidth]{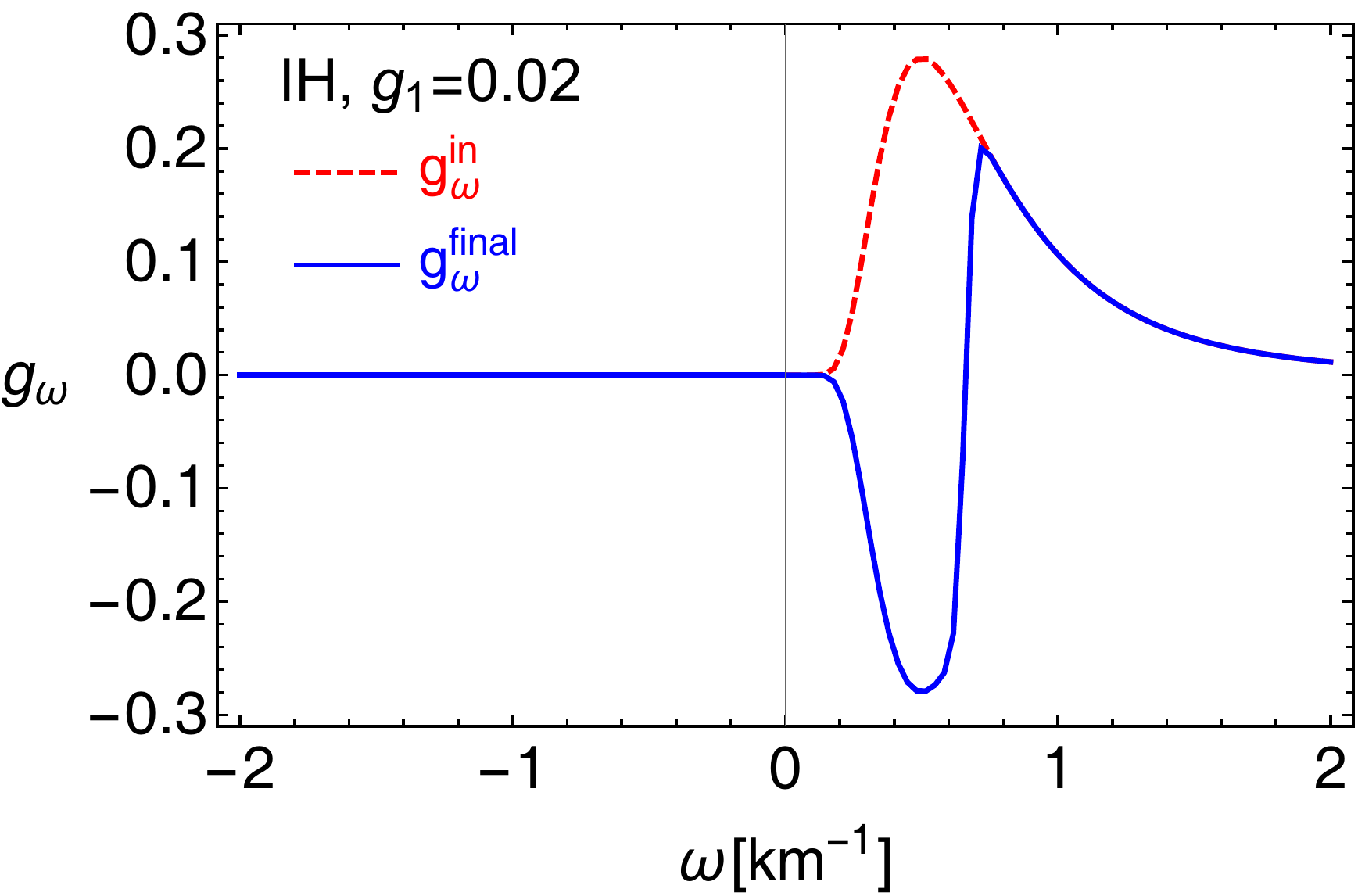}
 \,\,\,\,\,\,\,\includegraphics[width=0.484\textwidth]{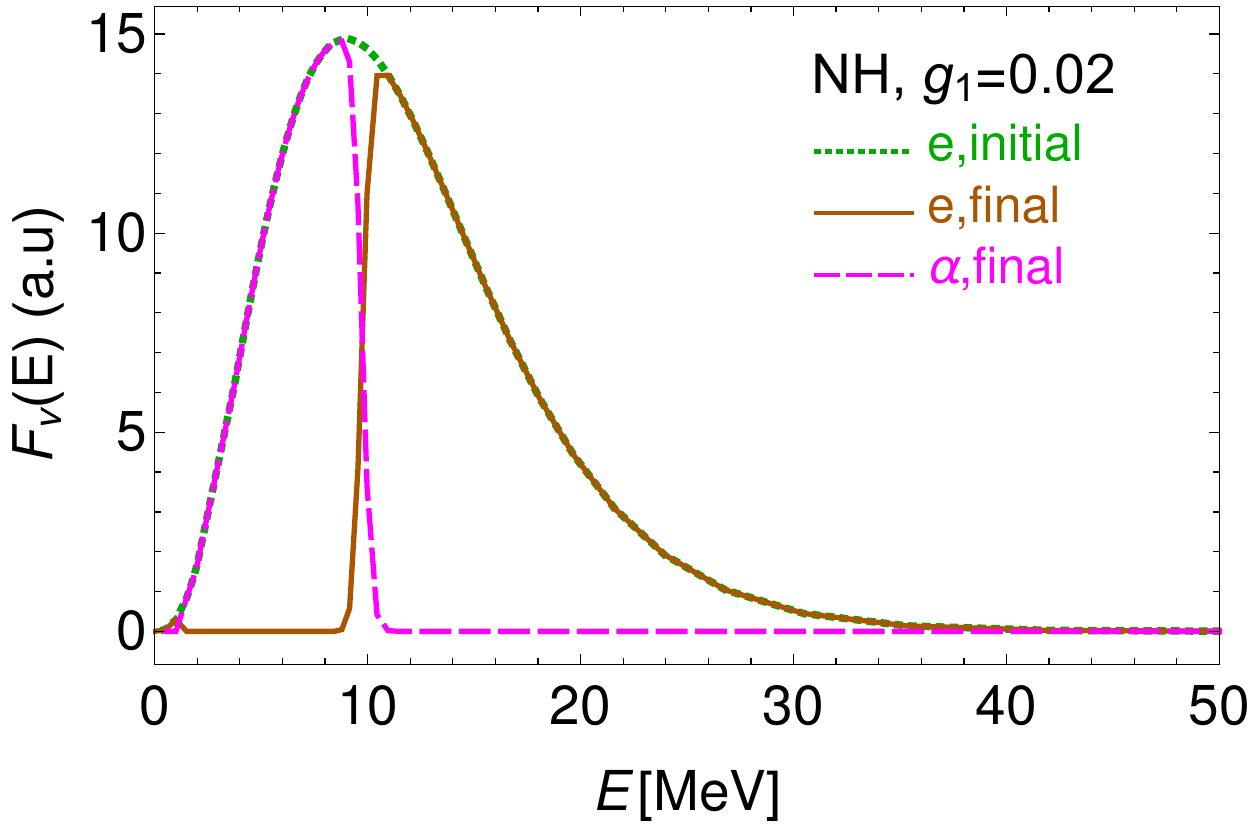}
 \,\,\,\,\,\,\,\,\,\includegraphics[width=0.483\textwidth]{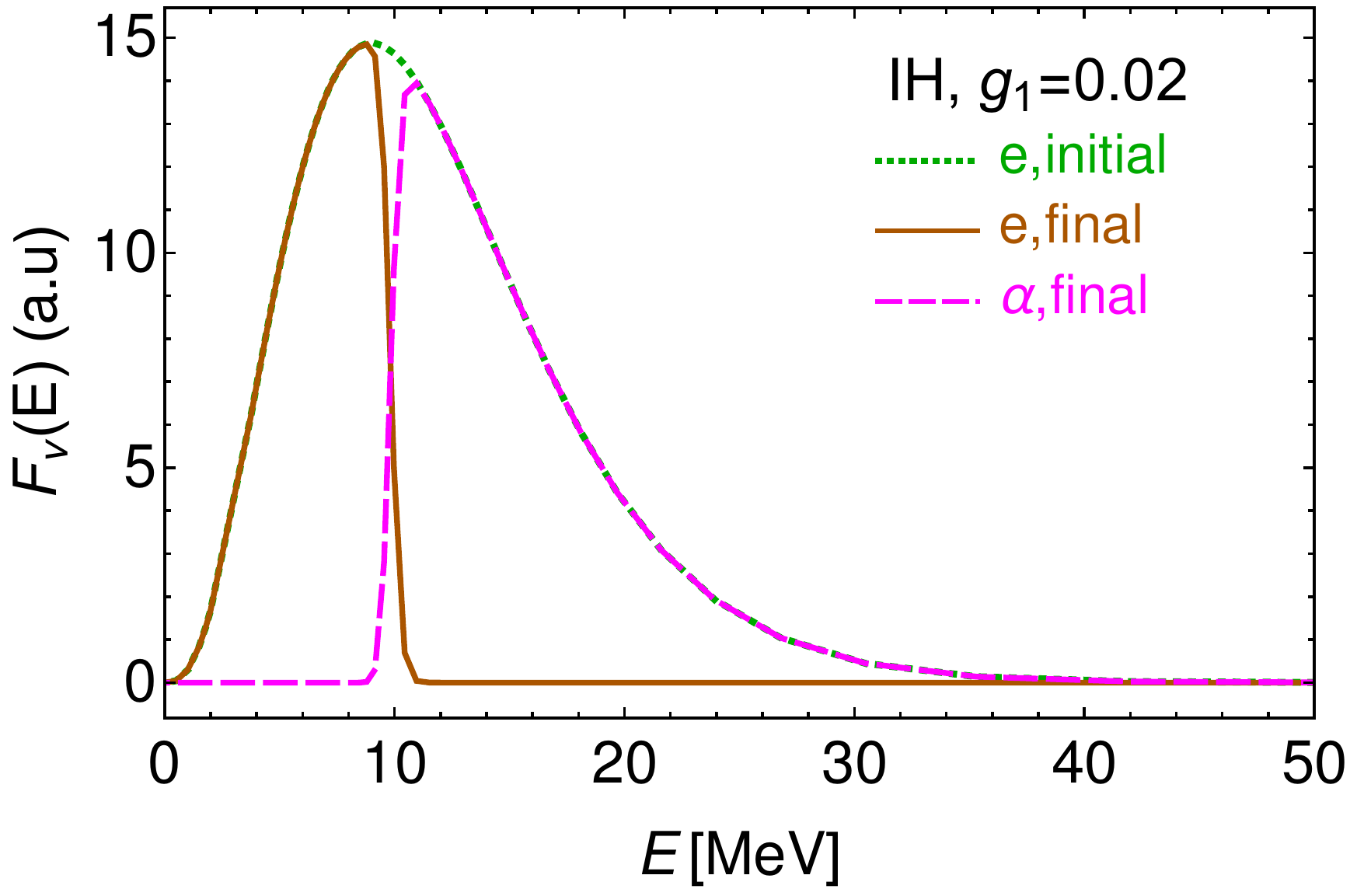} 
\caption{Effects of $\nu_e\leftrightarrow\nu_\alpha$ collective oscillations on an initial $\nu_e$ spectrum during the neutronization burst, for $g_1=0.02,~g_3=0$. Left panels: NH. Right panels: IH. We have taken $E=\Delta m^2_{\rm atm}/(2\w)$.}
\label{fig9}
\end{figure}

In order to see how the presence of FV-NSSI affects the neutrino spectrum during neutronization burst, we consider the following initial flux~\cite{Keil:2002in}
\begin{equation}
 F_{\nu}^0(E_\nu)\propto\frac{E_{\nu}^3}{ \langle E_{\nu}\rangle^4} e^{-4\frac{E_{\nu}}{\langle E_{\nu}\rangle}}\,.
\end{equation}
We choose the average energy $ \langle E_{\nu_e}\rangle= 12~{\rm MeV}$ \cite{2010A&A...517A..80F}. The fluxes for the $\bar{\nu}_e,\nu_\alpha$ and $\bar{\nu}_\alpha$  have all been taken to be zero during the neutronization epoch.
With these parameters, the initial and final $\nu_e$ spectra (just after the $\nu_e\leftrightarrow\nu_\alpha$ collective oscillations are over) are shown in Fig.\,\ref{fig9} for both NH (left panel) and IH (right panel).
As expected, we observe a distinct split in the $\nu_e$ spectrum in both hierarchies.
\begin{itemize}
 \item 
In NH, for high energies, the final spectrum  
is identical to the original $\nu_e$ spectrum, whereas at lower energies, all the $\nu_e$s get converted to $\nu_\alpha$s. 
As a result, the average energy of the $\nu_e$ spectrum will  increase and a sharp rise would be observed after a certain critical energy.

\item In IH, the exact opposite behaviour is observed. For low energies, the final spectrum  
is identical to the original $\nu_e$ spectrum, whereas at higher energies, all the $\nu_e$s get converted to $\nu_\alpha$s. 
As a result, the average energy of the $\nu_e$ spectrum will decrease and and its tail would be replaced by a sharp drop. 
\end{itemize}

The analysis so far takes into account only two-flavor collective conversions. This could be followed further by another collective flavor conversion, in the stepwise process In principle, collective conversions $\nu_e\leftrightarrow\nu_y$ and $\nu_e\leftrightarrow\nu_x$ may happen in a stepwise manner \cite{Dasgupta:2007ws, Dasgupta:2008cd, Choubey:2010up}, where
\begin{equation}
 \nu_y\equiv\cos\theta_{23}\,\nu_\mu+\sin\theta_{23}\,\nu_\tau\,,\qquad \nu_x\equiv-\sin\theta_{23}\,\nu_\mu+\cos\theta_{23}\,\nu_\tau\,.
\end{equation}
The spectra after these collective transformations will further be affected by MSW flavor conversions at the H and L-resonances \cite{Kuo:1989qe,Dighe:1999bi}. The final spectra arriving at the Earth would be 

\begin{eqnarray}
  F_{\nu_e}^{\rm NH}&=&  \bigg[|U_{e3}|^2\left(1-P_{ey}-P_{ex}\right)+|U_{e2}|^2 P_{ey}+|U_{e1}|^2 P_{ex}\bigg] F_{\nu_e}^0\,, \nonumber\\
  F_{\nu_e}^{\rm IH}&=&  \bigg[|U_{e2}|^2\left(1-P_{ey}-P_{ex}\right)+|U_{e3}|^2 P_{ey}+|U_{e1}|^2 P_{ex}\bigg] F_{\nu_e}^0\,,
\end{eqnarray}
where $P_{ey}$ and $P_{ex}$ are the collective flavor conversion probabilities for $\nu_e\leftrightarrow\nu_y$ and $\nu_e\leftrightarrow\nu_x$ respectively. We have taken the H and L-resonances to be adiabatic as is clear from the large values of the corresponding mixing angles \cite{Dighe:1999bi,Esteban:2016qun, nufit}.
In the absence of NSSI, $P_{ey}=P_{ex}=0$ and hence 
\begin{equation}
 F_{\nu_e}^{\rm NH}=|U_{e3}|^2\, F_{\nu_e}^0  \, , \qquad  F_{\nu_e}^{\rm IH}= |U_{e2}|^2 F_{\nu_e}^0 .
\end{equation}
Since $|U_{e3}|^2\simeq0.025\,,|U_{e2}|^2\simeq0.3$ and the flux during the neutronization burst is well-predicted \cite{Kachelriess:2004ds}, the two hierarchies can be distinguished by observing the number of events during the first $\sim 20$ ms of a SN neutrino signal.
 
\begin{figure}[!t]
  \includegraphics[width=0.47\textwidth, height=0.326\textwidth]{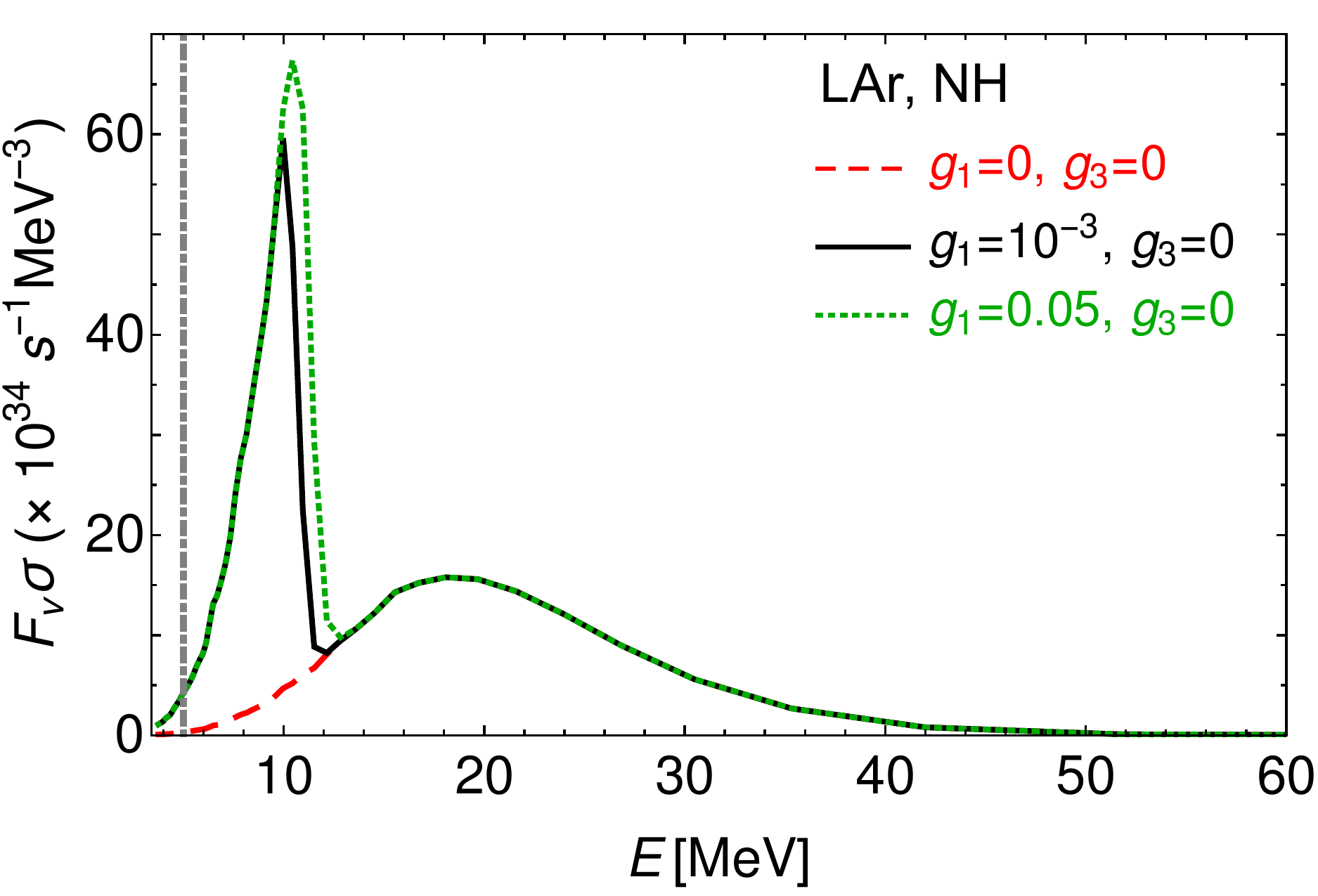}~
  \includegraphics[width=0.52\textwidth, height=0.317\textwidth]{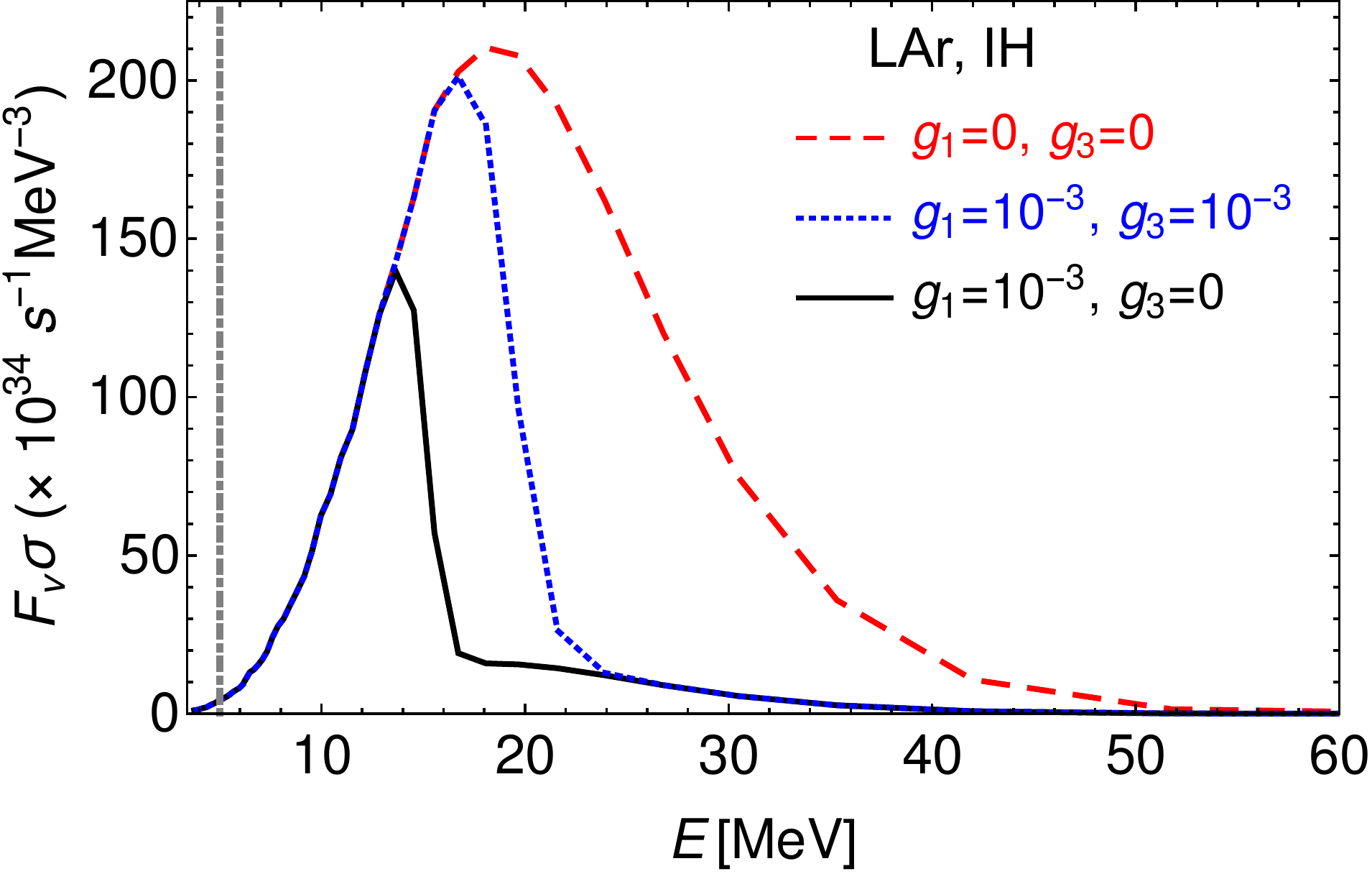}
\caption{The quantity (flux $\times$ cross-section) for charged-current events $(\nu_e + ^{\rm 40} Ar\rightarrow ^{\rm 40} K^* + e^-)$ in a liquid Argon detector in different scenarios.
 The cross-sections have been taken from \cite{Gil-Botella:2016sfi}. The threshold energy of the detector is taken to be $E=5\,{\rm MeV}$ \cite{Acciarri:2015uup}, as shown in grey dotdashed lines. Left panel: NH.  Right panel: IH.
 }
\label{fig10}
\end{figure}

This picture changes with the introduction of NSSI.  The non-zero values of $P_{ey}$ and $P_{ex}$, combined with the sharp energy dependent spectral split features in these quantities will affect the final number of events as well as the $\nu_e$ spectral shape. For an illustration, we choose
the scenario where $P_{ex}=0$ and show the quantity (flux $\times$ cross-section) at a liquid Argon (LAr) detector in Fig.\,\ref{fig10}. Note that the exact position of the spectral split would depend on the initial flux as well as the values of NSSI parameters. In the scenario shown in the figure, increasing the value of $g_1$ tends to shift the split to higher energies in NH and lower energies in IH. On the other hand, the presence of a non-zero $g_3$ would cause pinching of the swap. This corresponds to shifting the split to lower energies in NH and to higher energies in IH. 
If further $P_{ex}\neq0$, it could give rise to multiple spectral splits. Since such features can be present in both hierarchies, the identification of mass hierarchy from the neutronization burst \cite{Dighe:1999bi,Kachelriess:2004ds} would become difficult.

The observation of such a spectral split during the neutronization epoch would indicate the presence of NSSI.
This however would need a sufficiently large number of events and a very good resolution in time and energy to resolve these splits. With a neutrino flux of 
$\sim 10^{57}$ $\nu_e$s during the neutronization burst of a SN
 at $10\,{\rm kpc}$, one would expect up to $ \mathcal{O}(100)$ events in a $40\,{\rm kt}$ liquid Argon detector. In water Cherenkov detectors, where $\nu_e$ flux will be detected through the elastic scattering $\nu_e+e^-\rightarrow\nu_e+e^-$  and the energy determination is not so good, the signals of NSSI may be discerned if the expected number of events are observed to be too high for the NH scenario and too low for the IH scenario. 
 For the $500\,{\rm kt}$ Hyper-Kamiokande, one would expect up to $ \mathcal{O}(100)$ of events during this neutronization burst.

Note that the results in this paper are obtained under the single-angle approximation. Multi-angle effects and possible consequent effects of matter may modify the final spectra. However the distinctive effects of NSSI, in particular the formation of spectral splits where none would be present otherwise, could survive and are worth exploring further.

\section{Summary and Discussions }
\label{sec:5}
In this paper, we have investigated the effects of non-standard self-interactions (NSSI) on collective oscillations of supernova (SN) neutrinos, motivated by \cite{Blennow:2008er} and the rather weak limits on the NSSI couplings \cite{Bilenky:1999dn}.
Using a flavor-pendulum picture to get an analytical understanding, we have performed a comprehensive study of the impact of flavor-preserving (FP-NSSI) as well as flavor-violating NSSI (FV-NSSI) on the flavor evolution of neutrinos. We work with two neutrino flavors and in the single-angle approximation, and expect that the qualitative features of the effects of NSSI would be captured even in this simplified scenario. Indeed, many interesting results significantly distinct from the Standard Model (SM) expectations are seen to emerge with the addition of NSSI.

For an ensemble of neutrinos and antineutrinos of a fixed energy, we have shown that for large enough NSSI, the predictions for the two mass hierarchy interchange, i.e., 
flavor conversions can happen in NH, whereas they can vanish in IH. 
For a typical neutrino-neutrino potential in a SN, the FP-NSSI are observed to act like a matter term, causing a delay in the onset of flavor conversions. The FV-NSSI result in the violation of flavor lepton number, and hence do not preserve the initial neutrino-antineutrino flux asymmetry.

We have also analyzed the effects of NSSI on a box-spectrum of neutrinos and antineutrinos over a range of energy modes, in order to clarify how NSSI affects the spectral swaps.
In the presence of FP-NSSI, spectral swaps develop around a spectral crossing.
The FP-NSSI lead to the pinching of the spectral swaps, i.e., a decrease in their width and height. 
The flavor lepton number violation arising from FV-NSSI leads to interesting observations: while in the SM, the swaps have to develop around the zero crossing of the $g_\w$-spectra, the presence of FV-NSSI may cause swaps to appear away from spectral crossings, and even in the absence of spectral crossings. 

An important consequence of such a flavor lepton violation with FV-NSSI is the presence of collective oscillations during the neutronization burst epoch of a SN leading to low (high) energy conversion of $\nu_e$ to $\nu_\alpha$ in NH (IH). This would alter the neutronization burst signal. Using a realistic $\nu_e$ spectrum during the neutronization burst and taking into account the effect of MSW resonance inside the star, we demonstrate the presence of 
spectral splits in the final spectra. Since collective effects would otherwise be absent during this epoch, the presence of such splits can be a clear indication of NSSI. This could also make the identification of hierachy during neutronization burst harder.

Note that our work uses a two-neutrino framework and a single-angle approximation. Nonetheless, these simple approximations already bring out possible interesting features with the introduction of NSSI. A more realistic treatment will require a detailed three flavor study, with the inclusion of multi-angle effects and matter effects. 
The possible non-standard interactions among neutrinos and charged fermions may also give rise to further interesting features.
It might also be interesting to study how the presence of NSSI affects formation of the neutrino spectra and dynamics of the SN core, as suggested in~\cite{Amanik:2006ad}.
 Thus the introduction of NSSI to SN dynamics and neutrino flavor conversions can open up a plethora of new effects and rich phenomenology.

%%%%%%%%%%%%%%%%%%%%%%%%%%%%%%%%%%%%%%%%%%%%%%%%%%%%%%%%%%%%%%%%%%%%%%%%%%
\section*{Acknowledgments}
%%%%%%%%%%%%%%%%%%%%%%%%%%%%%%%%%%%%%%%%%%%%%%%%%%%%%%%%%%%%%%%%%%%%% 
We would like to thank Basudeb Dasgupta and Alessandro Mirizzi for insightful discussions, suggestions, and comments on the manuscript. We would also like to thank Steen Hannestad and Georg Raffelt for useful discussions. We thank the Max-Planck-Institut f\"ur Physik, Munich, and the Aarhus University, for hospitality during the initial days of this project. 
The work of A.\,Das and M.S. was supported by the Max-Planck Partnergroup ``Astroparticle Physics" of the Max-Planck-Gesellschaft awarded to Basudeb Dasgupta. This project has received partial support from the European Union's Horizon 2020 research and innovation programme under the Marie-Sklodowska-Curie grant agreement Nos. 674896 and 690575.

\end{document}